\documentclass[12pt]{article}
\usepackage{epsfig,cite}
\usepackage {amsmath}
\usepackage{amssymb}
\include{epsf}

\usepackage{times}

\def\beq{\begin{equation}}
\def\eeq{\end{equation}}
\def\bey{\begin{eqnarray}}
\def\eey{\end{eqnarray}}

\def\msun{M_\odot}

\def\lsim{\mathrel{\raise.3ex\hbox{$<$\kern-.75em\lower1ex\hbox{$\sim$}}}}
\def\gsim{\mathrel{\raise.3ex\hbox{$>$\kern-.75em\lower1ex\hbox{$\sim$}}}}
\def\Msun{M_\odot}

\begin{document}

\begin{center}
UTTG-09-10
\hspace{0.3cm} TCC-016-10
\hspace{0.3cm} MCTP-10-30
\hspace{0.3cm} FERMILAB-PUB-10-319-A-PPD
\end{center}
\vskip 0.2in
\begin{center}
{\large{\bf  Black Holes in our Galactic Halo: Compatibility with FGST and PAMELA Data and Constraints on the First Stars}}
\end{center}
\begin{center}
\vskip 0.2in
{\bf Pearl Sandick,}$^{1}$
{\bf Juerg Diemand,}$^2$
{\bf Katherine Freese,}$^{3}$
{\bf and Douglas Spolyar}$^{4,5}$

\vskip 0.1in

{\it $^1${Theory Group and Texas Cosmology Center, The University of Texas at Austin, TX 78712}\\}
{\it $^2${Institute for Theoretical Physics, University of Z\"urich, CH-8057, Switzerland}\\}
{\it $^3${Michigan Center for Theoretical Physics, University of Michigan, Ann Arbor, MI 48109}\\}
{\it $^4${Center for Particle Astrophysics, Fermi National Accelerator Laboratory, Batavia, IL  60510}\\}
{\it $^5${Department of Astronomy and Astrophysics, The University of Chicago, Chicago, IL 60637}\\}

\end{center}
\vskip 0.1in
\begin{abstract}

$10-10^5 \msun$ black holes with dark matter spikes that formed in 
early minihalos and still exist in our Milky Way Galaxy today are examined in light
of recent data from the Fermi Gamma-Ray Space Telescope (FGST). 
The dark matter spikes surrounding black holes in our Galaxy are sites of significant dark matter annihilation. We examine the signatures of annihilations into gamma-rays, $e^+/e^-$,
and neutrinos.  We find that some significant fraction of the point sources 
detected by FGST might be due to dark matter annihilation near black holes in our Galaxy.  We obtain limits on the properties of dark matter annihilations in the spikes using the information in the FGST First Source Catalog as well as the diffuse gamma-ray 
flux measured by FGST.
We determine the maximum fraction of 
high redshift minihalos that could have hosted the formation of the first generation of stars
and, subsequently, their black hole remnants.  The strength
of the limits depends on the choice of
annihilation channel and black hole mass; limits are strongest for the heaviest black holes and
annhilation to $b \bar{b}$ and $W^+W^-$ final states.   The larger black holes considered in this paper may arise as the 
remnants of Dark Stars after the dark matter fuel is exhausted and thermonuclear burning runs its course; thus FGST observations may be used to constrain the properties of Dark Stars.   Additionally, we comment on the excess positron flux found by PAMELA and its possible interpretation in terms of dark matter annihilation around these black hole spikes. 

\end{abstract}



\section{Introduction}

The very first generation of stars, known as Population III.1, likely formed from metal-free, molecular hydrogen-cooled gas at the center of dark matter minihalos of $\sim 10^6 \Msun$ at $z \gtrsim 10$~\cite{HTL1996,brommlarson,yoshida03,ripamontiabel,barkanaloeb,bcl 1999}. 
Simulations indicate that they were quite massive, 
typically more than 100 $\Msun$.   If dark matter is made of Weakly Interacting Massive Particles (WIMPs), which are typically their
own antiparticles, then the first phase of stellar evolution may be a Dark Star (DS) phase~\cite{spolyar08}, during which the star is powered by dark matter (DM) annihilations 
prior to nuclear fusion.
Indeed, if the DS phase persists for an extended period of time before fusion sets in, the star may grow to be correspondingly more massive: as long as dark matter annihilation powers the star, it remains cool enough to continue to accrete matter and grow ever larger. 
Initial work showed that they can grow to be $\sim 1000\Msun$~\cite{Freese:2008wh}, while more recent 
studies, taking into account triaxial halos with a variety of particle orbits, demonstrate that they may even become super-massive $\gtrsim 10^5 \Msun$~\cite{Freese:2010re}.  
Subsequent to the initial paper on Dark Stars \cite{spolyar08},
 other authors have explored the repercussions of DM heating
in the first stars, including~\cite{iocco1, FSA2008, sellwood, Taoso, Yoon, ripamontione, iocco2, Spolyar:2009nt, schleichera, schleicherb, ripamontietal, umeda, Sivertsson, zackrisson1, zackrisson2}.

When the DM fuel inside a DS runs out, it collapses and heats up to become a standard fusion-powered star.  
If some Population III.1 stars experience a DS phase, the Initial Mass Function (IMF) for fusion-powered stars may be quite different than what might otherwise be expected~\cite{tan08}.  Specifically, the IMF would be determined by the length of the DS phase, i.e. by how long it took each DS to exhaust its DM fuel supply.
The ``dark power'' may have lasted different amounts of time in 
different DSs: millions to billions of years for some stars or, in extreme cases, even until today.  
Stars in the mass range $140-260\Msun$  would have ended their
lives as pair instability supernovae, leaving no remnants~\cite{hegerwoosley, stardeath}, but stars outside this
mass range would collapse
to black holes.  It is the latter case that we consider in this paper.

Our work, though motivated by the DS scenario, applies generally to black holes at the centers of minihalos, regardless of their origin.
Many black holes of mass $10-10^5 M_\odot$ that formed at the centers of minihalos survive in the universe today.
Assuming some fraction of high redshift minihalos hosted Population III.1 star formation,
one can estimate the distribution of their remnant black holes today.  Each remnant black hole will be surrounded by a region of enhanced dark matter density, which we refer to as a ``spike.''
Although some of the 
original minihalos would have merged with other DM halos, resulting in disruption,
one can still follow the evolution of the black holes and their DM spikes in simulations.
Here, we use the Via Lactea-II simulation to track black hole spikes from the redshift of their formation to $z=0$.
In this way, we estimate the black hole population in a galaxy like our Milky Way today.  

When the baryonic gas in a minihalo collapses to form a structure, be it a star or black hole, dark matter is dragged in, creating a DM spike around the central object.  
The  star or black hole provides a gravitational potential that causes the surrounding
 DM to be pulled inward, into and around the object.  The resulting enhanced DM density may be computed via adiabatic contraction,
 assuming that the DM orbital timescale is shorter than the timescale of the changing gravitational potential.  
As the star reaches the end of its lifetime, collapsing to become a black hole, the DM spike remains in place.
The enhanced DM density in these spikes leads to an enhanced rate of DM annihilation,
 which scales as the square of the DM density. Unstable DM annihilation products undergo hadronization and decay, with typical final products being gamma-rays, $e^+/e^-$ pairs, and neutrinos, each of which may provide a significant signal at a variety of ground- and/or space-based detectors.

In this paper, we explore the detection prospects for gamma rays produced in dark matter annihilations in the DM spikes surrounding black holes for a range of star formation scenarios, black hole masses, and dark matter annihilation modes.  The current data from the Fermi Gamma-Ray Space Telescope (FGST) are used in a two-pronged approach to constrain the number of black holes in the Milky Way halo and, consequently, the number of DSs that could have formed at early times.  We use the FGST First Source Catalog~\cite{fgstFSC} to find the minimal distance to the nearest DM spike such that it is not brighter than the brightest source observed by FGST. From the predicted distribution of such spikes in the Milky Way halo, we extract a limit on the fraction of minihalos in the early universe to host a black hole (and survive as a DM spike in our Galactic halo today).  We also use the FGST measurement of the diffuse gamma-ray background~\cite{fgstEGB} to constrain the population of DM spikes contributing to the diffuse flux today, thereby setting a second limit on the fraction of minihalos in the early universe to become DM spikes in our Galactic halo.

Several groups have previously 
studied signatures from DM annihilation in DM overdensities, or spikes, around black holes. 
Gondolo and Silk~\cite{GS} first pointed out the possibility of a DM spike around the $\sim 10^6 \Msun$ Super-Massive Black Hole (SMBH) at the center of the Galaxy; they coined the terminology ``spike'' to differentiate the gravitationally contracted DM due to the existence of the black hole from the more standard possibility of cusps at the centers of DM halos in cold dark matter scenarios of structure formation.  The  enhancement in the dark matter density in the spike due to the SMBH at the center of the Milky Way was later shown to be likely reduced by a variety of effects including mergers, formation of the initial black hole away from the Galactic Center, and gravitational scattering with stars~\cite{ulliokam, preto, bertonemerritt, Bertone:2005xv}.  Zhao and Silk \cite{ZS} suggested DM spikes around Intermediate Mass Black Holes, remnants of Pop.~III stars, similar to the work in this paper.  Shortly thereafter, Bertone, Zentner, and Silk (BZS)~\cite{bzs}, with follow-up work in~\cite{bftz},~\cite{tabp}, and~\cite{bls}, 
studied formation and evolution histories for black holes 
and proposed looking for signatures of DM annihilation in the spikes around them in the Galaxy today.  They chose two scenarios to examine: (i) 100 $\Msun$ black holes left over as remants from the first stars, and (ii) $10^5 \Msun$ black holes that formed in larger $10^7 \Msun$ halos due to accretion disks.  They too estimated the distribution of unmerged spikes that reside in the Milky Way today for each scenario.  Our work, although similar to theirs, differs quantitatively in the following  respects.  First, we identify star-forming DM halos in a different way, as described in Section~\ref{sec:sfr}. Second, since we are motivated by a potentially long-lived DS phase for Pop.~III.1 stars, we consider a variety of black hole masses between $10$ and $10^5 \Msun$.  Third, we use
recently released data from FGST to constrain the current population of black holes and, consequently, the initial population of their progenitors.  In particular, we consider data from both the FGST First Source Catalog~\cite{fgstFSC} and the measurement of the diffuse gamma-ray flux~\cite{fgstEGB}.

In Section~\ref{sec:sfr} we discuss the remnant black hole population in the Milky Way halo, and in Section~\ref{sec:spikes} we discuss the DM spikes surrounding those black holes.  In Section~\ref{sec:signal} we present the signal from dark matter annihilation in DM spikes in the Milky Way halo, that due to spikes that appear as point sources and the expected diffuse gamma-ray flux. We then use FGST point source data and the measurement of the diffuse gamma-ray background to place constraints on the star formation history in a variety of dark matter annihilation models, as presented in Section~\ref{sec:fDS}. In Section~\ref{sec:epnu} we briefly address signatures in positrons and neutrinos and discuss prospects for future work. Finally, in Section~\ref{sec:conclusions} we present our conclusions.

\section{Remnant Black Hole Population in the Milky Way Halo}
\label{sec:sfr}

In order to estimate the observable consequences of annihilation in the DM spikes around black holes, we need to estimate the number of black holes in the Milky Way halo today that are the remnants of
the first stars.  Ideally, we would like to know the 
Initial Mass Function and Star Formation Rate of Pop.~III.1 stars;
however, what guidance we can get from simulations is limited by our understanding of feedback processes in the early universe.
During the DS phase, since the stars are extremely cool there is very little
ionizing radiation and therefore very little feedback.  However, as soon as the first fusion-powered stars form, they produce ionizing photons which
influence and hinder the formation of stars in neighboring regions.  
The effects of feedback on star and/or protogalaxy formation have been studied in Refs.~\cite{Haiman,AhnShapiro,greifbromm2006,WiseAbel,MckeeTan,SakumaSusa}, among others.
Because of the many uncertain aspects of theoretical studies of Population III.1 star formation, in this paper
we follow the method discussed in the remainder of this section.

We use the Via Lactea-II simulation to estimate the number and mass distribution of DM minihalos as a function of redshift.  The minihalos 
that hosted Pop.~III.1 stars must have had masses in a very particular range at the time of star formation.
Pop.~III.1 stars could only form when a cooling mechanism for the collapse of the baryonic cloud arose. The first accessible mechanism to cool the gas was via excitations of molecular hydrogen, the fraction of which present in a minihalo is related to the temperature, which in turn can be written in terms of the mass and redshift of the minihalo. We use the parametrization of Ref.~\cite{Hcooling} for the minimum halo mass in which star formation could occur:
\beq
M_{min}^{halo} \approx 1.54 \times 10^5 \Msun \Big(\frac{1+z}{31}\Big)^{-2.074}.
\label{eq:Hcooling}
\eeq
Because of the hierarchical nature of structure formation, there are far more smaller halos than larger ones.  Our results are therefore not sensitive to the maximum halo mass for Pop.~III.1 star formation, which we take to be $M_{max}^{halo}=10^7 \Msun$.
See also \cite{AhnShapiro}.

Next we assume that some fraction, $f_{DS}^0$, of viable minihalos hosted a Pop III.1 star (e.g. a DS). 
We also assume
 that each  star ended its life in collapse to a black hole.  
 Neglecting, for now, black hole mergers, the comoving number density of black holes as a function of redshift is then
\beq
N_{BH}(z) = f_{DS}^0 \, N_{halo}(z),
\eeq
where $N_{halo}(z)$ is the comoving number density of minihalos in which Population III.1 star formation was possible.

The duration of the DS phase is highly variable, and DSs may live vastly different amounts of time before running out of DM fuel, becoming fusion-powered, and finally collapsing to a black hole.
However, the lifetime of the star is not important as long as it has reached the black hole stage today.
We also note that we do not consider
further accretion of mass onto the black hole; this is the most conservative assumption regarding potential detectability and constraints on DS scenarios.

At some time between the beginning of Population III.1 star formation and the end of reionization at $z \sim 6$, hydrogen deuteride (HD) cooling would have become possible in minihalos, at which point massive Population III.1 (and DS) formation would have given way to less-massive Population III.2 star formation~\cite{greifbromm2006}, however there are few constraints on when this transition occured.  Motivated by recent discussions of the effect of feedback from the very first stars on subsequent star formation and reionization~\cite{AhnShapiro,greifbromm2006,WiseAbel,MckeeTan,SakumaSusa,WiseAbel2}, we consider three scenarios for the termination of Population III.1 star formation at redshift $z_f$.  These scenarios are hereafter noted as Early, Intermediate, and Late, following Ref.~\cite{greifbromm2006}, with $z_f \approx 23$, $15$, and $11$, respectively, as shown in Table~\ref{tab:Zfinal}.
For each case, we assume that Pop.~III.1 star formation was only possible in minihalos with masses in the range discussed above and prior to the termination redshift.

\begin{table}[h!]
\begin{center}
\begin{tabular}{|c|c|c|}
\hline
Label &  $z_f$ & Representation \\
\hline
Early & 23 & green  \\
Intermediate & 15 & red \\
Late & 11 & blue \\
\hline
\end{tabular}
\caption{Three scenarios for the redshift at which Population III.1 star formation ceases, $z_f$.  In the third column we list the color used to represent each case in the following figures.
\label{tab:Zfinal}}
\end{center}
\end{table}

We determine the $z=0$ distribution of DM spikes throughout the Galactic
halo from the Via Lactea II (VL-II) cosmological N-body simulation~\cite{vl2}. With a particle mass of $4.1 \times 10^3 M_\odot$,
VL-II is the first simulation of the Galactic DM halo that is able
to directly resolve the small progenitors (minihalos) which are the
formation sites of Population III.1 stars.

As discussed above, since the truncation
redshift for Pop.~III.1 star formation is poorly constrained,  we consider three different scenarios: Early
($z_f=23.1$), Intermediate ($z_f=14.8$), and Late ($z_f=11.2$). At these redshifts we identify all minihalos
between the minimum mass in Eq.~\ref{eq:Hcooling} and a maximum mass of
$10^7 M_\odot$ \footnote{Using a larger maximum mass of $10^8 M_\odot$ gives
virtually the same results, a consequence of the steepness of the halo mass functions 
for these redshifts and mass ranges.} and assign one Population III.1 star
to its most bound particle (tracer particle). We assume that this
Population III.1 star will then form a black hole surrounded by a DM spike, as described in Section~\ref{sec:spikes}, and that
both the black hole and the spike will survive until $z=0$. The $z=0$ position
of the tracer particles directly relates to the distribution of DM
spikes in this scenario. The simulation naturally includes the effects
of dynamical friction and tidal stripping as the minihalos fall into
larger host halos. Eventually many of these minihalos fall below $10^5 M_\odot$ 
and are no longer resolved in VL-II. From this point on the
remaining minihalo with its black hole and DM spike is represented as a point
mass in the simulation, i.e. the dynamical friction of the stripped
minihalo against the Galactic DM halo is neglected, but this effect is
very small for objects below $10^5 M_\odot$ even near the Galactic center~\cite{binneytremaine}.

We mention here two caveats: First, VL-II does not include the
formation of the Galaxy which could contract the DM halos and
therefore also the spike distribution. However, contraction may be
much smaller than the classical adiabatic models assume, if it exists
at all (e.g. Ref.~\cite{dutton} argues for DM halo expansion
during galaxy formation). Second, when two minihalos (each containing
a black hole) merge, it is possible that the black holes form a close
binary which would destroy the DM spikes. Our model does not account
for that. However, we estimate, as described in detail below, that
mergers may change the number of DM spikes in our Galaxy today by at most a 
factor of 2 for the highest black hole masses considered, and very little for 
low mass black holes.

The limited time and spatial resolution of VL-II and the
lack of massive black hole particles in this DM-only simulation make it
impossible to identify black hole mergers directly. To estimate the fraction
of DM spikes which could have been destroyed in mergers we simply
identify for each spike its nearest companion black hole in 24 snapshots between $z=11$ 
and $z=0$~\footnote{A list of the
exact reshifts is available at www.physik.uzh.ch/$\sim$diemand/vl.}.
For simplicity we assume here that the spikes orbit around each
other on a circular orbit within an isothermal sphere, in this case
the dynamical friction time is simply~\cite{binneytremaine} 
\begin{equation}\label{tfr}
t_{\rm fric} = \frac{19\, \rm Gyr}{\ln \Lambda}
\left( \frac{\Delta r}{0.5\, \rm kpc} \right)^2
\frac{\Delta v}{20\, \rm km s^{-1}}
\frac{10^5 \Msun}{M_{\rm BH+DMspike}} \; ,
\end{equation}
where $\Delta r$ and $\Delta v$ are the distance and relative
velocities of the two neighbouring black holes. The Coulomb logarithm is set
to $\ln \Lambda = 6$. Equation~\ref{tfr} is evaluated for all DM
spikes at each of the 24 time steps between $z=11$ and $z=0$ and all DM spikes which
might have merged by today according to their $t_{\rm fric}$ are added
up. The fraction of DM spikes potentially lost through mergers is
significant only for the most massive black holes and for $f_{DS}^0$ close to
one: For the Late scenario, with $f_{DS}^0=1$ and $M_{\rm BH+DMspike} =
10^5 \Msun$ the fraction of possibly lost spikes is $f_{merged}=0.49$. Lowering the
mass to $10^4 \Msun$ results in $f_{merged}= 0.24$, and going down to $10^3
\Msun$ gives $f_{merged}=0.076$. These estimates illustrate that these mergers would result in disruption of, at most, fewer than half of all Milky Way black holes for the largest black hole masses considered here, and much smaller fractions for lower black hole masses.
Since it is not possible to directly identify black hole mergers in VL-II, for the following analysis we define $f_{DS}$ to be the fraction of surviving black holes with surrounding DM spikes, such that it is related to the initial fraction of star-forming minihalos, $f_{DS}^0$, as
\beq
f_{DS} = f_{DS}^0 (1-f_{merged}),
\eeq  
with the fraction of DM spikes destroyed in mergers, $f_{merged}$, itself a function of $f_{DS}^0$.  For small $f_{DS}^0$, mergers become very rare and $f_{DS} \rightarrow f_{DS}^0$.

\begin{figure}[h]
\begin{center}
\mbox{\epsfig{file=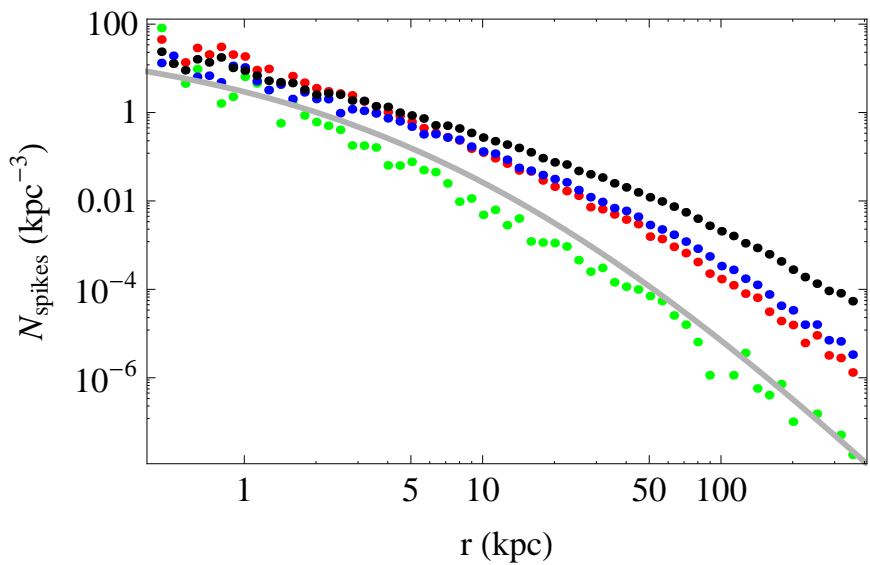,width=.7\textwidth}}
\end{center}
\caption{\it The number density of black hole spikes in the Milky Way as a function of Galactic radius for star formation models with Early (green), Intermediate (red), and Late (blue) $z_f$ as described in the text and for $f_{DS}=1$.  Curves have been 
obtained using the VL-II N-body simulation as described in the text.  The black points illustrate the
total dark matter density profile at $z=0$ in VL-II.  Also shown as a solid grey curve is the analytical fit found in Ref.~\cite{tabp,bzs}.  Our simulations show 409, 7983, and 12416 DM spikes in the Milky Way for the Early, Intermediate, and Late scenarios, respectively, assuming $f_{DS}=1$.
\label{fig:spikedists}}
\end{figure}

In Fig.~\ref{fig:spikedists} we show the number densities of DM spikes inside the Milky Way halo
as functions of Galactic radius for Early, Intermediate, and Late $z_f$. In the Early case, Pop.~III.1 star formation terminates at $z\approx 23$, so there were the fewest stars, and therefore the fewest black holes and surviving density spikes today. In the Intermediate and Late cases, Population III.1 star formation turns off at redshifts of roughly 15 and 11, respectively.  
For comparison, the total dark matter density profile at $z=0$ in VL-II is also shown; although the normalization of these points is
arbitrary, it is useful to illustrate that the total DM profile is more extended than the distribution of  black holes with DM spikes.

We can contrast our approach for finding the relevant DM minihalos to that of Ref.~\cite{bzs}.  At $z=18$,
they populated halos that constituted $3\sigma$ peaks in the smoothed primordial density field with seed black holes of initial mass $100 \Msun$.  
Using an analytical model of halo evolution, they simulated 200 statistical realizations of the
growth of a Milky Way-sized halo.
A fit to their resulting distribution of DM spikes is shown as the grey curve in Fig.~\ref{fig:spikedists}. 
Instead, we use a single iteration of the VL-II cosmological simulation and follow potential Population III.1 star-forming minihalos within the halo mass range and redshift ranges discussed above. 

Using their approach, Ref.~\cite{bzs} found that  $1027 \pm 84$ unmerged IMBHs and surrounding DM spikes are expected to exist in the Milky Way halo today.  Our simulations yield 409, 7983, and 12416 DM spikes in the Early, Intermediate, and Late scenarios, respectively, assuming $f_{DS}=1$. 
However, not every viable minihalo must have hosted a Pop.~III.1 star, in which case the total number of DM spikes in our Galactic halo would be roughly proportional to $f_{DS}^0$.  Again, for small $f_{DS}^0$, the fraction of DM spikes destroyed in mergers becomes negligible, and $f_{DS} \approx f_{DS}^0$.


\section{Dark Matter Density Spikes}
\label{sec:spikes}

The density profile of a dark matter spike surrounding a black hole is determined by adiabatic contraction of the dark matter halo around the central mass.  For concreteness, as our starting point we take the Navarro, Frenk, and White (NFW) profiles for both the baryons (15\% of the mass) and dark matter (85\% of the mass):
\beq \label{iniprofile}
\rho(r)=\frac{\rho_0}{r/r_s(1+r/r_s)^2},
\end{equation}
where $\rho_0$ is the scale density and $r_s$ is the scale radius~\cite{nfw}.
The scale density, $\rho_0$, can be re-expressed in terms of the critical density of the universe at a given redshift, $\rho_c(z)$, via
\beq
\rho_0=\rho_c(z)\frac{200}{3}\frac{C^3}{ln(1+C)-C/(C+1)},
\end{equation}
where we take the value of the concentration parameter to be $C\equiv r_{vir}/r_s = 3.5$, and $r_{vir}$ is the virial radius of the halo.  We find that decreasing the concentration parameter to $C=2$ results in a $\sim 30 \%$ decrease in the luminosity of each spike.  Sensitivity of DS properties to the concentration parameter is discussed in Ref.~\cite{ilie2010}.

We assume a flat $\Lambda$CDM universe with current matter density 
$\Omega_m=0.24$ and dark energy density $\Omega_\Lambda=0.76$.

We allow a point mass to grow at the center of the halo, 
mimicking the formation of a star and the subsequent remnant black hole.  To model the response of the
DM to this density growth we have used the simple Blumenthal {\it et al.} prescription for adiabatic contraction~\cite{blum}.  
It has been shown that this method obtains a DM density profile that is accurate to within a factor of 2~\cite{sellwood}.
The result of the adiabatic contraction is a roughly power law density profile. 

\begin{figure}[h]
\begin{center}
\mbox{\epsfig{file=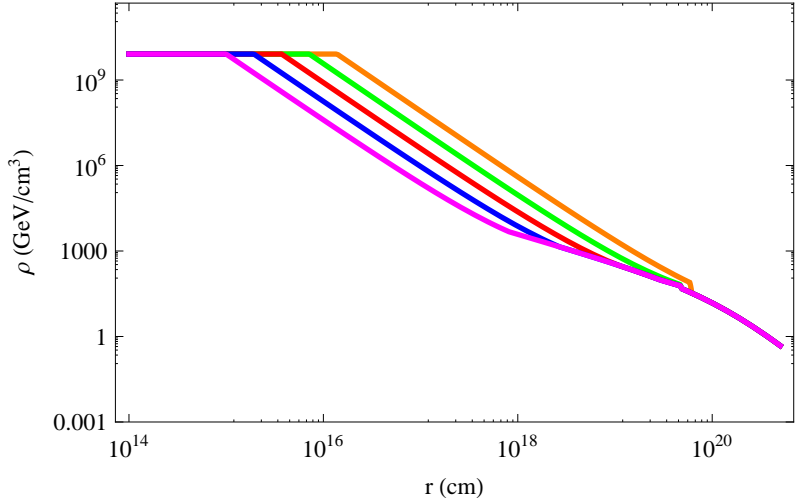,height=7cm}}
\end{center}
\caption{\it Density profiles for contracted dark matter halos surrounding black holes of mass $10 \Msun$ (magenta), $10^2 \Msun$ (blue), $10^3 \Msun$ (red), $10^4 \Msun$ (green), and $10^5 \Msun$ (orange), from bottom to top, assuming $m_\chi = 100$ GeV and that the central black hole in each case formed at $z=15$ in a halo of mass $10^6 M_\odot$. Note that $1\, \rm{pc} = 3.1 \times 10^{18}\, \rm{cm}$.
\label{fig:densityprofs}}
\end{figure}

In Fig.~\ref{fig:densityprofs} we show the contracted halo profiles today for DM spikes due to black holes of various masses for the case
where the central object formed at $z=15$. 
 We note that the power law portion of the profile is independent of WIMP mass.
  In the central regions, closest to the black hole, some of the DM has annihilated away in the time since the formation of the central mass.  We follow BZS who found an upper limit to the DM density
\beq
\label{eq:rhomax}
\rho_{max} = \frac{m_{\chi}}{\langle \sigma v \rangle t_{BH}},
\eeq
where $t_{BH}$ is the lifetime of the 
central mass,
roughly $1.3 \times 10^{10}$ years for a star that formed at $z=15$. 
Thus we take the density profile in the inner region
to be a flat plateau with this density out to the radius $r_*$, defined by $\rho (r_*)=\rho_{max}$, 
beyond which the density profile follows that expected from adiabatic contraction of the minihalo.
We note, however, that a proper treatment of this inner region is more complicated: Because dark matter particles in triaxial halos may follow box or chaotic orbits~\cite{valluri}, 
some of the DM particles that are expected to have annihilated with each other in the center of the spike would not have been on orbits that would have returned them to the center anyway.
Thus, for triaxial halos, it is harder to deplete the density in the central region.  
The use of Eq. \ref{eq:rhomax} may therefore underestimate the amount of DM remaining in the center, but on these long timescales it is an acceptable estimate.

\begin{figure}[h]
\begin{center}
\mbox{\epsfig{file=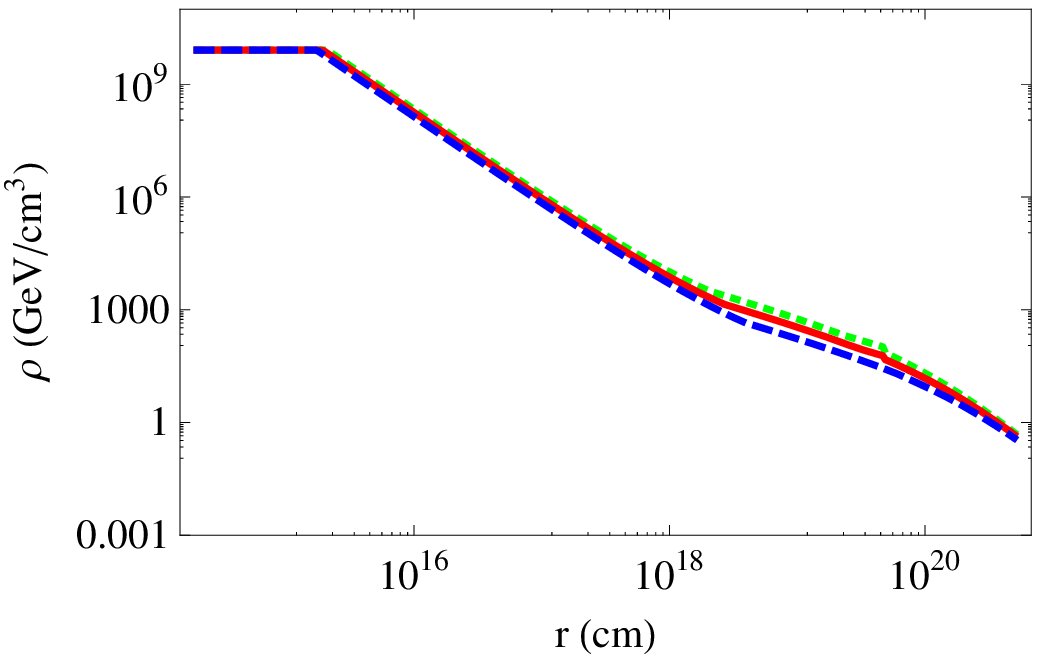,width=.45\textwidth}}
\mbox{\epsfig{file=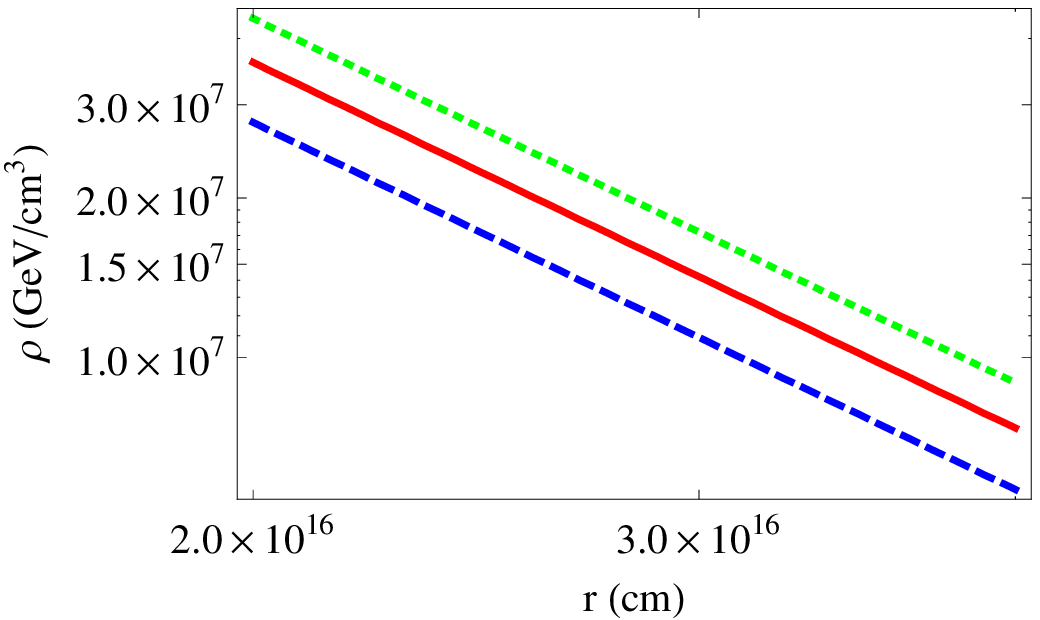,width=.45\textwidth,height=4.8cm}}
\end{center}
\caption{\it Density profiles for contracted dark matter halos surrounding black holes of mass $10^2 \Msun$ for black hole formation at $z=10$ (blue, dashed), $z=15$ (red, solid), and $z=20$ (green, dotted). The initial halo mass is taken to be $10^6 M_\odot$ for each of the three cases, and $C=3.5$. The right panel zooms in to a range of radii that contribute significantly to the total DM annihilation rate in the spike to show more clearly the differences between the three cases.  Note that $1\, \rm{pc} = 3.1 \times 10^{18}\, \rm{cm}$.
\label{fig:halocomp}}
\end{figure}

Fig.~\ref{fig:halocomp} illustrates the dependence of the DM density profile on the redshift of formation of the central DS and subsequent black hole.
The right panel 
  reveals that there is nearly a factor of two difference between the density profiles for  black holes that form at $z=10$ and $z=20$ at the radii shown,
  which correspond to regions that contribute significantly to the total DM annihilation rate 
  (we note that only a fraction of a percent of the total annihilations in each spike occur at $r \gtrsim 1$ pc).  The reason for this 
  difference is the fact that at higher redshifts, densities are higher as $(1+z)^3$ and the halos have smaller radii (n.b. we keep $C$ 
  fixed throughout).
 Thus, we expect differences in the flux of annihilation products from DM spikes created at different redshifts, all other factors being equal.

\section{Gamma Ray Signal from Dark Matter Annihilations}
\label{sec:signal}

For a Majorana dark matter particle with mass $m_\chi$ and annihilation cross section times velocity $\langle \sigma v \rangle$, the rate of WIMP annihilations in a DM spike is
\beq
\Gamma = \frac{\langle \sigma v \rangle}{2 m_\chi^2}\int_{r_{min}}^{r_{max}} dr \, 4 \pi r^2 \, \rho_{DM}^2,
\label{eq:rate}
\eeq
with $r_{min}$ and $r_{max}$ defining the volume of the DM spike in which annihilations occur. At a minimum, $r_{min}$ should be equal to the Schwarzschild radius of the black hole, $r_{Sch}=2 G m_{BH}$ for a black hole of mass $m_{BH}$ and Newton's constant, $G$. We take $r_{min}=4 r_{Sch}$, though our results are not sensitive to this choice for any $r_{min} \lesssim 10^{14}$ cm, corresponding to $\sim 3 \times 10^{(8-i)}$ Schwarzschild radii for central black hole masses of $10^i M_\odot$.

We choose as a benchmark scenario $\langle \sigma v \rangle = 3 \times 10^{-26}$ cm$^3$s$^{-1}$, in agreement with the measured dark matter abundance today for thermal WIMP dark matter, and consider several WIMP candidates, defined by mass and annihilation channel.  Calculations are performed for WIMP masses of 100, 200, 500, 1000, and 2000 GeV and Standard Model final states $b \bar{b}$, $W^+W^-$, $\tau^+ \tau^-$, and $\mu^+ \mu^-$. 
The resulting spectrum of photons from annihilation to final state $f$, $dN_f/dE$, is computed with PYTHIA~\cite{pythia}.  
For $\chi \chi \rightarrow \mu^+\mu^-$, the photon spectrum comes from final state radiation and is given by
\beq
\frac{dN_{\mu^+\mu^-}}{dx} = \frac{\alpha}{\pi} \left(\frac{x^2-2x+2}{x}\right) 
\left[ \ln \left(\frac{s(1-x)}{m_\mu^2}\right)-1\right],
\eeq
where $x\equiv E_\gamma/m_\chi$, the center-of-mass energy squared is $s=4m_\chi^2$, and $\alpha \approx 1/137$~\cite{fsr}. We note that WIMP candidates such as neutralinos typically annihilate to a variety of final states, $f$, with branching fractions $B_f$ such that the total annihilation cross section $\langle \sigma v \rangle$ can be expressed as the sum over final states
\beq
\langle \sigma v \rangle = \sum_f \langle \sigma v\rangle_f = \langle \sigma v \rangle \sum_f B_f,
\eeq
with the final sum evaluating to unity.  Consequently, the rate of annihilations to final state $f$ in a DM spike may be expressed as
\beq
\Gamma_f= B_f \Gamma.
\label{eq:rate2}
\eeq

\begin{figure}[h!]
\begin{center}
\mbox{\epsfig{file=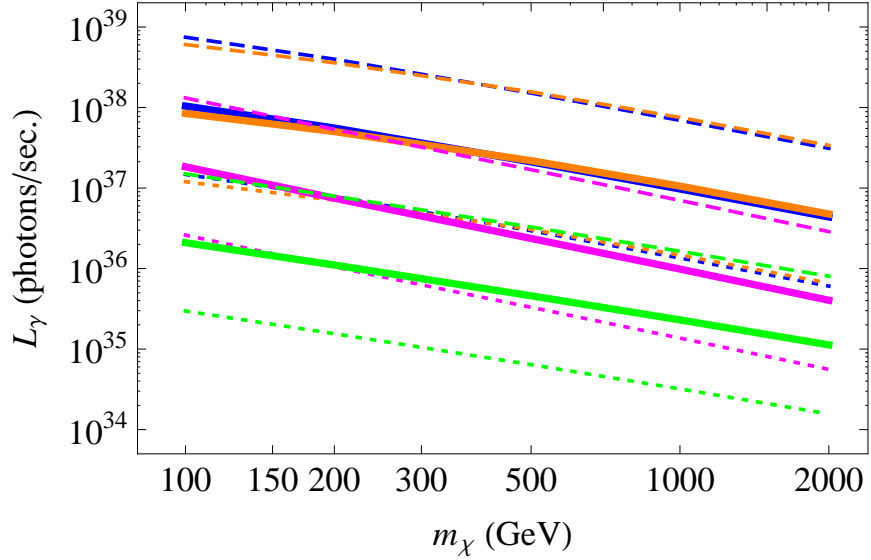,width=.7\textwidth}}
\end{center}
\caption{\it The intrinsic luminosity of photons with energy above 1 GeV from a single DM spike as a function of WIMP mass in the Intermediate $z_f$ scenario. The thick curves are for a central black hole of $10^2 M_\odot$ and the thinner dotted and dashed curves are for central black holes of $10 M_\odot$ and $10^3 M_\odot$. From top to bottom, the blue, orange, magenta, and green contours represent final states $b \bar{b}$, $W^+W^-$, $\tau^+\tau^-$, and $\mu^+\mu^-$, respectively.
\label{fig:IntLum}}
\end{figure}

The intrinsic photon luminosity from dark matter annihilations in any dark matter spike is then
\beq
{\cal L} = \int dE \,\sum_f \frac{dN_f}{dE} \,\Gamma_f,
\eeq 
with $\Gamma_f$ given by Equations~\ref{eq:rate} and~\ref{eq:rate2}. This quantity is plotted as a function of WIMP mass in Fig.~\ref{fig:IntLum} assuming $B_f=1$ for each of the final states $b\bar{b}$ (blue), $W^+W^-$ (orange), $\tau^+\tau^-$ (magenta), and $\mu^+\mu^-$ (green) and for DM spikes surrounding black holes of mass $10 M_\odot$ (thin dotted curves), $100 M_\odot$ (thick solid curves), and $1000 M_\odot$ (thin dashed curves).  We see that increasing the mass of the central black hole by an order of magnitude increases the luminosity by nearly an order of magnitude for all WIMP masses and final states.  Also, heavier WIMPs result in significantly lower luminosity, as there are fewer of them in each DM spike and therefore the annihilation rate is lower.  We note also the similarity of the luminosities for final states $b\bar{b}$ and $W^+W^-$, a consequence of the similarity of the photon spectra $dN/dE$ in these two cases. 

\begin{table}[h!]
\begin{center}
\begin{tabular}{|l|c|c||l|c|c|}
\hline
Model &  Mass (GeV) & Final State & Model &  Mass (GeV) & Final State \\
\hline 
b100    &   100    &  $b \bar{b}$ & $\tau$100   &   100   &  $\tau^+ \tau^-$ \\
b1T     &   1000   &  $b \bar{b}$ & $\tau$1T   &   1000   &  $\tau^+ \tau^-$ \\
W100    &   100    &  $W^+ W^-$  & $\mu$100    &   100   &  $\mu^+ \mu^-$ \\
W1T     &   1000   &  $W^+ W^-$  & $\mu$1T    &   1000   &  $\mu^+ \mu^-$ \\
\hline
\end{tabular}
\caption{example WIMP annihilation models
\label{tab:wimpmodels}}
\end{center}
\end{table}

For simplicity, in the following sections, we focus on 8 combinations of WIMP mass and final state, which we label with one particle of the final state and the WIMP mass, as shown in Table~\ref{tab:wimpmodels}.  Furthermore, we assume that dark matter has only one dominant anihilation mode, and consider only $B_f=1$ for each final state.  If dark matter annihilations result in more than one final state, as for standard neutralino or Kaluza-Klein dark matter, our results must be scaled appropriately.  
Given a WIMP annihilation model, the distribution of DM spikes in the Milky Way halo (as shown in Fig.~\ref{fig:spikedists}), and the characteristic density profile of each DM spike (for example, as shown in Figs.~\ref{fig:densityprofs} and~\ref{fig:halocomp}), we calculate the expected gamma-ray signal.  In Section~\ref{sec:PS} we discuss FGST point sources, and in Section~\ref{sec:diffuse} we discuss the diffuse gamma-ray flux.


\subsection{Point Sources}
\label{sec:PS}

The differential flux of neutral particles from annihilations to final state $f$ in a DM spike with radius $r_{max}$ located some distance $D$ from our Solar System is given by
\beq
\frac{d\Phi_f}{dE}=\frac{\Gamma_f}{4 \pi D^2} \frac{dN_f}{dE},
\eeq
for $D \gg r_{max}$.  If $D$ is not large compared to $r_{max}$, however, an integral must be performed along the line-of-sight, $s$, and over the solid angle of interest on the sky, defined by the polar angle, $\theta$:
\beq
\frac{d\Phi_f}{dE}=\frac{B_f \langle\sigma v \rangle}{2 m_\chi^2} \frac{dN_f}{dE} \int_0^{\theta_{max}} d\theta \,2 \pi \sin\theta\int_0^{s_{max}} ds \, \rho_{DM}^2(r),
\eeq
where $r=\sqrt{D^2+s^2-2sD \cos\theta}$, $s_{max} > D+r_{max}$, and $\theta_{max}$ defines the solid angle to which the detector is sensitive as $\Delta\Omega = 2 \pi (1-\cos\theta_{max})$.

FGST has compiled a catalog of point sources identified at $\gtrsim 4 \sigma$ significance within the first year of data collection~\cite{fgstFSC}.  If a single DM spike is a sufficiently bright and compact source of gamma-rays, it may have been identified as a point source and recorded in the FGST First Source Catalog, which contains 1451 point sources, including 630 that are not
associated with sources in other astronomical catalogs.  Here 
we consider only point sources that are more than 10$^\circ$ away from the Galactic
plane (that is, $|b| > 10^\circ$), as it is expected that the region close to the plane will contain the majority of baryonic gamma-ray sources (pulsars, supernova remnants,
X-ray binaries, etc.).  Of the unassociated point sources in the FGST First Source Catalog, 368 have been detected
with greater than $5 \sigma$ significance and are more than 10$^\circ$ away from the Galactic plane.

Ref.~\cite{buckleyhooper} identifies a set of criteria for objects which should be identified as point sources by FGST in the first year of observation, which we adopt here, as well\footnote{In order to identify point sources in the first year data, the Fermi Collaboration have used a complicated diffuse emmission model and likelihood Test Statistic as detailed in Ref.~\cite{fgstFSC}. The sensitivity of our results to our choice of simplistic criteria is discussed in Secs.~\ref{sec:diffuse} and~\ref{sec:fDS}.}.  Specifically, $>50$ events per year must be observed by FGST and $>95\%$ of the events must come from within a cone of half-angle $2^\circ$ centered on the source.
A bound on our black hole sources can be placed as follows:  spikes bright enough to be identified as point sources must not be brighter than the brightest source in the the FGST First Source Catalog, which has a flux of photons with 100 MeV 
$< E_\gamma <$ 100 GeV of $1.25 \times 10^{-6}$ cm$^{-2}$s$^{-1}$.  Requiring that the flux not exceed this value establishes a minimal distance, $D_{min}^{PS}$, beyond which the spike must be located in order not to be brighter than any source in the FGST catalog.  

The DM spikes from adiabatically contracted minihalos around black holes have quite steep density profiles, so in all models considered we find that all sources located at distances $\geq D_{min}^{PS}$ are highly localized such that the $2^\circ$ requirement is more than satisfied. We would like to emphasize that the dark matter spikes in our
paper are indeed point-like objects, with the dark matter annihilation
signal coming from the spike rather than from the extended minihalo around
it. Of all the scenarios considered, the closest a spike can be to our
solar system is ~0.01 kpc (see Fig. 5).  Even in this case, $> 99\%$ of the
signal comes from within a cone of half-angle $0.01^{\circ}$ (that is, most
of the annihilations occur within $\sim.002$ pc of the central black hole).
Since the 95\% confidence level Fermi containment angle is at least an
order of magnitude larger than this at all energies~\cite{FermiPSF}, this nearest spike is
decidedly a point-like object.

Before proceeding, we would like to point out that the brightest FGST point source is actually identified as associated with the Vela pulsar.  The brightest source that is {\it unassociated} with known sources in other wavelengths has a flux of $5.78\times 10^{-8}$ cm$^{-2}$s$^{-1}$.  The factor of $\sim22$ difference between the flux from the Vela source and that from the brightest unassociated source means that had we chosen to base $D_{min}^{PS}$ on the unassociated source we would find each $D_{min}^{PS}$ to be a factor of $\sqrt{22}=4.7$ larger. We will return to this point in the following analysis.

\begin{figure}[h!]
\begin{center}
\mbox{\epsfig{file=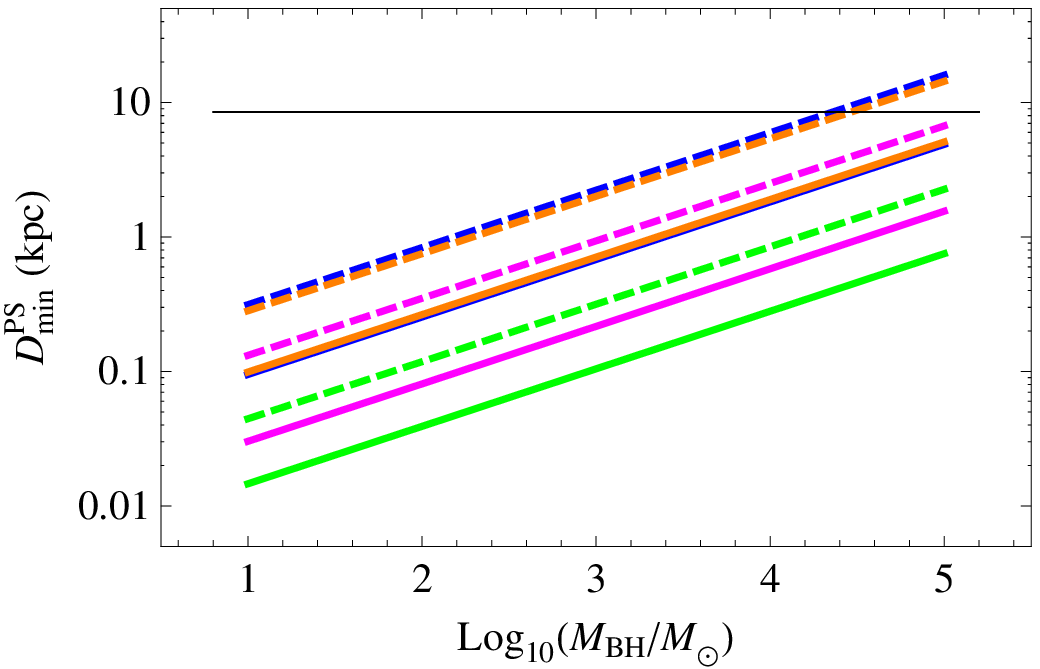,width=.48\textwidth}}
\mbox{\epsfig{file=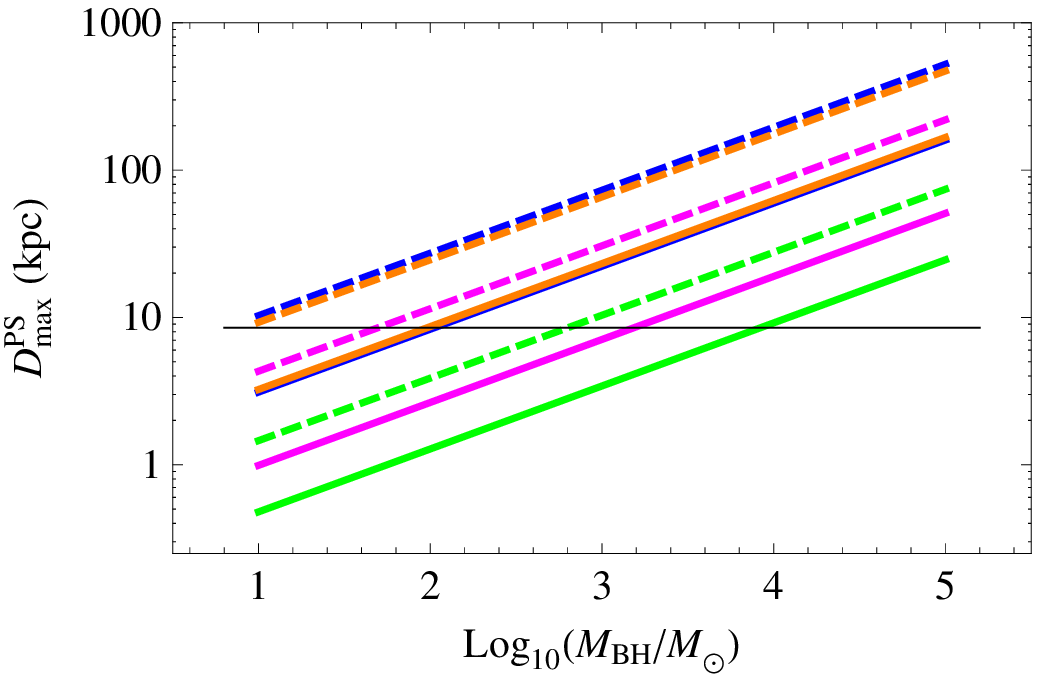,width=.48\textwidth}}
\caption{In the left panel, we display the minimal distance from our Solar System of a single spike such that it does not exceed the flux of the brightest source in the FGST First Source Catalog as a function of central black hole mass for Intermediate $z_f$.  From top to bottom, the contours are for models b100 (blue dashed), W100 (orange dashed), $\tau$100 (magenta dashed), b1T (blue solid), W1T (orange solid), $\mu$100 (green dashed), $\tau$1T (magenta), and $\mu$1T (green solid).  In the right panel, we show the maximal distance from our Solar System of a single spike such that it would appear as a $\gtrsim 5\sigma$ point source to Fermi for the same cases. We assume that $B_f=1$ for each of the annihilation modes considered.  If the total annihilation cross section is the sum of branching fractions to different Standard Model final states, $D_{min}^{PS}$ and $D_{max}^{PS}$ would be scaled by $\sqrt{B_f}$. In each panel, the horizontal black line indicates the distance to the Galactic center.
\label{fig:PointSourceDist}}
\end{center}
\end{figure}

In Fig.~\ref{fig:PointSourceDist}, we display in the left panel the minimal distance, $D_{min}^{PS}$, at which a point source must be located in order not to exceed the largest flux from any point source measured by FGST for the eight example cases in Table~\ref{tab:wimpmodels}.  Also plotted in Fig.~\ref{fig:PointSourceDist} is a horizontal line indicating the distance from our Solar System to the Galactic center. Note that for 100 GeV WIMPs annihilating to $b\bar{b}$ or $W^+W^-$, if $m_{BH} \gtrsim 2\times 10^4 M_\odot$ the nearest spike to our Solar System must be at least as distant as the Galactic center. From the spike distributions in Fig.~\ref{fig:spikedists}, we see that this is extremely unlikely.  In fact, within only 5 kpc of our Solar System, we expect there to be $\sim 7$ DM spikes for Early termination of star formation, and 94 DM spikes in both the Intermediate and Late cases, assuming $f_{DS}=1$.  Since the number of spikes in any volume scales with $f_{DS}$, we find that $f_{DS}$ can be quite constrained for very large central black holes.  
However, for most of the DM models shown, if the central black hole is $\lesssim 10^3 M_\odot$, spikes may be located within 1 kpc of our Solar System. No matter which of the three choices for $z_f$ we adopt, we expect that less than one DM spike would be located within 1 kpc of our Solar System, even for $f_{DS}=1$.  Again, had we chosen to base our analysis on the brightest unassociated point source, all curves in the left panel of Fig.~\ref{fig:PointSourceDist} would be shifted up by a factor of $\sim4.7$.

In the right panel of Fig.~\ref{fig:PointSourceDist}, we show the maximal distance, $D_{max}^{PS}$, at which a single DM spike would be identified by FGST as a $\gtrsim 5\sigma$ point source.  These distances were determined by requiring 50 events per year in FGST from each spike.  For 100 GeV WIMPs annihilating to $b\bar{b}$ or $W^+W^-$ around large black holes, as in the upper right portion of the right panel of Fig.~\ref{fig:PointSourceDist}, all DM spikes in the Milky Way halo would be identified as point sources by FGST. In the opposite extreme, for annihilation of 1 TeV WIMPs to $\tau^+\tau^-$ or $\mu^+\mu^-$ in the DM spike surrounding a 10 $M_\odot$ black hole, any spike further than $\sim1$ kpc would not be bright enough to be identified as a point source. As mentioned above, it is not guaranteed that there are any DM spikes within 1 kpc of our Solar System; thus  no DM spikes may be bright enough to appear as FGST point sources.
The results shown in Fig.~\ref{fig:PointSourceDist} are for Intermediate $z_f$ (star formation ending at $z_f \approx 15$). In the Early and Late $z_f$ scenarios there is a $\sim15\%$ decrease and a $\sim13\%$ increase in $D_{max}^{PS}$, respectively, and similar changes in $D_{min}^{PS}$.

In Table~\ref{tab:nPS} we list the number of DM spikes that would have been identified as $5\sigma$ point sources by FGST in the first year of operation in each of the WIMP annihilation scenarios in Table~\ref{tab:wimpmodels} for five black hole masses and Early, Intermediate and Late $z_f$. The number of such sources with Galactic latitude $|b| > 10^\circ$ is shown in parentheses next to each entry.  There are 368 unassociated $5\sigma$ point sources at $|b|>10^\circ$ in the FGST First Source Catalog~\cite{fgstFSC,buckleyhooper}.  One can see that many models studied here greatly overproduce FGST point sources if $f_{DS}=1$, but recall that the number of point sources scales with $f_{DS}$.  We note also that not all sources with the requisite brightness have been detected by FGST, for example due to local background fluctuations or spectral shapes that did not pass all analysis cuts.  Even with these considerations, it is clear from Table~\ref{tab:nPS} that many conclusions regarding $f_{DS}$ are highly model-dependent.  One can say with certainty, however, that $f_{DS}$ is most constrained for large central black holes in the Intermediate and Late $z_f$ cases, and possibly entirely unconstrained for smaller central black holes and if Pop.~III.1 star formation terminated at $z_f \gtrsim 20$. 

Additionally, we remind the reader that $D_{min}^{PS}$ has been determined based on the brightness of the brightest source observed by FGST, rather than by the brightness of the brightest {\it unassociated} point source, thereby over-estimating the number of nearby point sources in Table~\ref{tab:nPS}. 
If we assume that all gamma-rays from associated sources are in fact due to the objects with which they are associated (and not DM spikes), then any DM spike must not be brighter than the brightest unassociated source. $D_{min}^{PS}$ would therefore increase by a factor of 4.7 and consequently the number of point sources between $D_{min}^{PS}$ and $D_{max}^{PS}$ would decrease.   

Furthermore, as we have defined $D_{max}^{PS}$ based on the assumption that objects that generate 50 events per year in FGST would be identifiable as point sources, if the number of required events to pick out a point source is actually larger than this, then the numbers of point sources in Table~\ref{tab:nPS} are, again, too large.  
In fact, the diffuse gamma-ray background is not entirely uniform, and the Fermi Collaboration's ability to identify point sources in the data depends on complex modelling of the diffuse gamma-ray sky and the location of the sources in the sky~\cite{fgstFSC}.  Since the number of spikes that would appear as point sources depends on the flux required for each spike to be identified as a point source, and since this flux depends on the region of the sky in which the spike is located, we do not use our estimates of the numbers of spikes in our Galactic halo to constrain $f_{DS}$. We encourage the reader to interpret the numbers in Table~\ref{tab:nPS} merely as reasonable estimates, with the actual number of point sources likely being somewhat smaller.

\begin{table}[h!]
\begin{center}
\begin{tabular}{|c|c c c | c c c|}
 \multicolumn{1}{c}{} & \multicolumn{3}{c}{\bf{b100}} & \multicolumn{3}{c}{\bf{b1T}} \\
\hline
$m_{BH}/M_\odot$ & Early & Int. & Late & Early & Int. & Late \\
\hline
10 & 195 (70) & 1117 (649) & 557 (387)
   &  $\sim 2$ ($\sim 2$) & 21 (17) & 14 (11) \\ 
$10^2$ &  304 (151) & 3247 (2263) & 2935 (2186)
       &  147 (45) & 586 (372) & 281 (215)  \\
$10^3$ &  380 (213) & 5715 (4283) & 6754 (5305) 
       & 284 (135) & 2788 (1895) & 2340 (1708)   \\
$10^4$ &   381 (217) & 7237 (5548) & 10608 (8486)
       &   372 (207) & 5213 (3870) & 5866 (4575) \\
$10^5$ &   158 (128) & 5918 (4831) & 10946 (8946)
       &   392 (224) & 7069 (5402) & 9998 (7980) \\
\hline\noalign{\bigskip}
\multicolumn{1}{c}{} & \multicolumn{3}{c}{\bf{W100}} & \multicolumn{3}{c}{\bf{W1T}} \\
\hline
$m_{BH}/M_\odot$ & Early & Int. & Late & Early & Int. & Late \\
\hline
10 & 176 (58) & 868 (504) & 392 (289)
   &   $\sim 2$ ($\sim2$) & 24 (20) & 16 (13)    \\ 
$10^2$ & 294 (144) & 3011 (2073) & 2618 (1930)
       &   159 (50) & 692 (420) & 321 (242)  \\
$10^3$ & 377 (211) & 5461 (4072) & 6279 (4915)
       & 288 (139) & 2881 (1969) & 2457 (1801)  \\
$10^4$ & 388 (222) & 7152 (5473) & 10318 (8242)
       &  374 (209) & 5320 (3957) & 6043 (4718)   \\
$10^5$ & 169 (136) & 6140 (4998) & 11181 (9123)
       &   391 (223) & 7107 (5434) & 10144 (8099)  \\
\hline\noalign{\bigskip}
\multicolumn{1}{c}{} & \multicolumn{3}{c}{\bf{$\tau$100}} & \multicolumn{3}{c}{\bf{$\tau$1T}}  \\
\hline
$m_{BH}/M_\odot$ & Early & Int. & Late & Early & Int. & Late \\
\hline
10 &  $\sim 7$ ($\sim 5$) & 59 (47) & 36 (30)
   & $\ll 1$ ($\ll 1$) & $\lesssim 1$ ($\lesssim 1$) & $\lesssim 1$ ($\lesssim 1$)  \\ 
$10^2$ &  211 (81) & 1360 (808) & 768 (518)
       & $\sim 1$ ($\sim 1$) & 13 (11) & $\sim 8$ ($\sim 7$)  \\
$10^3$ &   314 (159) & 3515 (2480) & 3312 (2492)
       &   52 (27) & 311 (225) & 169 (133)  \\
$10^4$ &   381 (214) & 5963 (4488) & 7302 (5757)
       &  269 (123) & 2439 (1618) & 1906 (1366) \\
$10^5$ &  368 (211) & 7258 (5575) & 10879 (8713)
       &   360 (197) & 4813 (3541) & 5191 (4023)   \\
\hline\noalign{\bigskip}
\multicolumn{1}{c}{} & \multicolumn{3}{c}{\bf{$\mu$100}} & \multicolumn{3}{c}{\bf{$\mu$1T}} \\
\hline
$m_{BH}/M_\odot$ & Early & Int. & Late & Early & Int. & Late \\
\hline
10 & $< 1$ ($<1$) & $\sim 2$ ($\sim 2$) & $\sim1$ ($\sim1$)
   & $\ll 1$ ($\ll 1$) & $\ll 1$ ($\ll 1$) & $\ll 1$ ($\ll 1$)  \\ 
$10^2$ & $\sim 5$ ($\sim 4$) & 42 (34) & 26 (22)
       &  $<1$ ($<1$) & $\sim 1$ ($\sim 1$) & $\sim1$ ($\lesssim 1$)  \\
$10^3$ &  195 (69) & 1132 (658) & 578 (400)
       &   $\sim 3$ ($\sim2$) & 28 (23) & 18 (15)  \\
$10^4$ &  305 (152) & 3278 (2288) & 2987 (2229)
       &   172 (56) & 846 (493) & 390 (287)   \\
$10^5$ & 380 (214) & 5752 (4314) & 6836 (5374)
       &  294 (143) & 3013 (2074) & 2629 (1939)   \\
\hline\noalign{\bigskip}
\hline
\end{tabular}
\caption{Number of FGST $5 \sigma$ point sources in several WIMP annihilation scenarios for each of the five black hole masses and Early, Intermediate, and Late $z_f$, assuming both $f_{DS}=1$ and $B_f = 1$. In parentheses are the number of such sources with 
Galactic latitude $|b| > 10^\circ$, to be compared with the 368 5$\sigma$ point sources with $|b|>10^\circ$ in the FGST First Source Catalog.
\label{tab:nPS}}
\end{center}
\end{table}


\subsection{Diffuse Flux}
\label{sec:diffuse}

Many of the DM spikes in the Milky Way halo may be too faint to be identified as point sources by FGST.  Indeed, even DM spikes that have large enough fluxes to be identified as point sources may have been missed as a result of selection effects; for example, if the local background near the spike is large~\cite{fgstSources}.  Those DM spikes that cannot be identified as point
sources may contribute to the diffuse flux. 
Based on the point source criteria in Sec.~\ref{sec:PS} and the simulated distribution of Pop.~III.1/DS remnant spikes in the Milky Way halo from VL-II, we calculate the contribution to the diffuse gamma ray flux from all faint spikes and make comparisons to the diffuse flux measured by FGST~\cite{fgstEGB}.

Since we use the diffuse flux to place limits on the population of black holes with DM spikes in the Milky Way halo, it is important that we not overestimate the diffuse flux. 
In other words, we are careful to exclude from the analysis any spikes that could be identified as
point sources in the one year FGST data. 
In Section~\ref{sec:PS}, we discuss the maximum distance $D_{max}^{PS}$ for point sources detectable by FGST.  Any black hole spikes located at distances less than $D_{max}^{PS}$ are excluded from our estimate of the diffuse flux. 
In fact, in the following analysis we take the more conservative approach of requiring an even larger minimal distance for spikes that contribute to the diffuse flux, $D_{min}^{diff.}$, as discussed below.

Ref.~\cite{buckleyhooper} estimate that the diffuse gamma-ray background produces 20 events per year per square degree above 1 GeV in FGST for $|b| > 60^\circ$, with a larger rate at lower Galactic latitudes.  
Here, we impose the condition that only spikes producing fewer than 20 events per year in FGST are included in the calculation of the diffuse gamma-ray flux.
As mentioned above, we assume that a 5$\sigma$ 
detection of a point source would require roughly 50 events. Since the number of events
scales inversely with distance squared, we estimate that the minimal distance at which a DM spike would likely contribute to the diffuse flux is 
\beq
D_{min}^{diff.}=\sqrt{\frac{50}{20}} \,D_{max}^{PS} \approx 1.6 \,D_{max}^{PS}.
\label{eq:dmindiff}
\eeq
We note that sources located farther from our Solar System than $D_{max}^{PS}$ and closer than $D_{min}^{diff.}$ are included neither as point sources nor as part of the diffuse gamma-ray flux.

In the following analysis, we assume that $D_{max}^{PS}$ and $D_{min}^{diff.}$ are independent of Galactic latitude.  However, in actuality, the background to detection of point sources does depend on Galactic latitude, and therefore both the flux required to generate a $5\sigma$ excess and the flux below which the source would be too dim to be identified as a point source should also depend, to some degree, on the Galactic latitude.
To be specific, at high Galactic latitudes the diffuse gamma-ray flux is lowest, so we expect that by choosing $D_{min}^{diff.}$ as in Eq.~\ref{eq:dmindiff}, we may slightly overestimate the contribution of high $|b|$ DM spikes to the diffuse gamma-ray flux.  Sources resulting in 20 events per year above 1 GeV in FGST for $|b|>60^\circ$ represent a $\sim 3\sigma$ fluctuation on the diffuse background, and therefore such sources may be bright enough to have been identified as point sources. 
However, Ref.~\cite{fgstSources} find that $16(\pm1.8)\%$ (with a systematic uncertainty of 10\%) of the GeV isotropic diffuse background is due to unresolved point sources, whose fluxes are large enough such that they could have been identified as point sources but have evaded detection due to selection effects.  For sources at Galactic latitudes in the range $10^\circ<|b|<20^\circ$, 20 events per year above 1 GeV is only a $\sim 2\sigma$ fluctuation on the background.

Given the assumption of a Galactic latitude-independent $D_{min}^{diff.}$ as explained above, we compute both the energy spectrum of the contribution to the diffuse gamma-ray background from all DM spikes at distances greater than $D_{min}^{diff.}$ and the angular distribution in the sky of the gamma-ray flux from those spikes.
The latter quantity is for reference only, and may be sensitive to our choice of universal $D_{min}^{diff.}$.  Had we allowed $D_{min}^{diff.}$ to vary with Galactic latitude, more spikes at low Galactic latitudes and fewer spikes at high Galactic latitudes would contribute to the diffuse flux, potentially resulting in a   
more anisotropic angular distribution than what has been calculated here.
One should therefore view the anisotropies of the angular distributions presented here as minimal.

In addition to neglecting the dependence on the Galactic latitude of the diffuse gamma-ray flux, we have also neglected any local deviations from the average value of the diffuse flux.  In fact, the diffuse gamma-ray background is not entirely uniform, and local deviations from the average affect the brightness required of a point source for it to have been identified on top of the local background.  We do not consider these issues here, and take instead a single value of $D_{min}^{PS}$ for each model, applied to all regions of the sky.

\begin{figure}[h!]
\begin{center}
\mbox{\epsfig{file=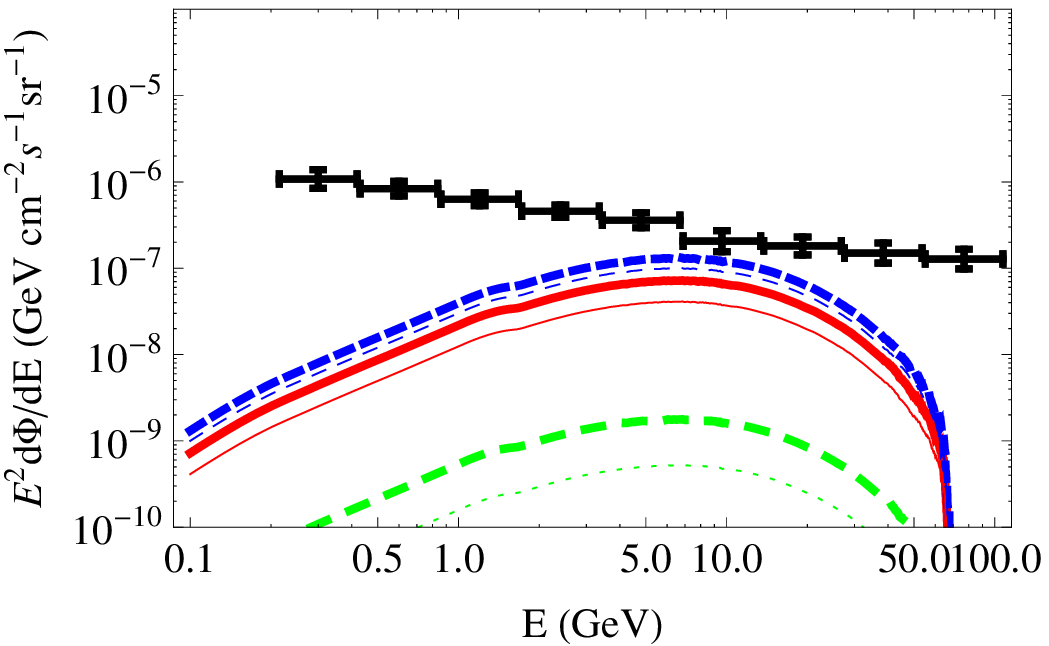,width=.4\textwidth}}\hspace{10mm}
\mbox{\epsfig{file=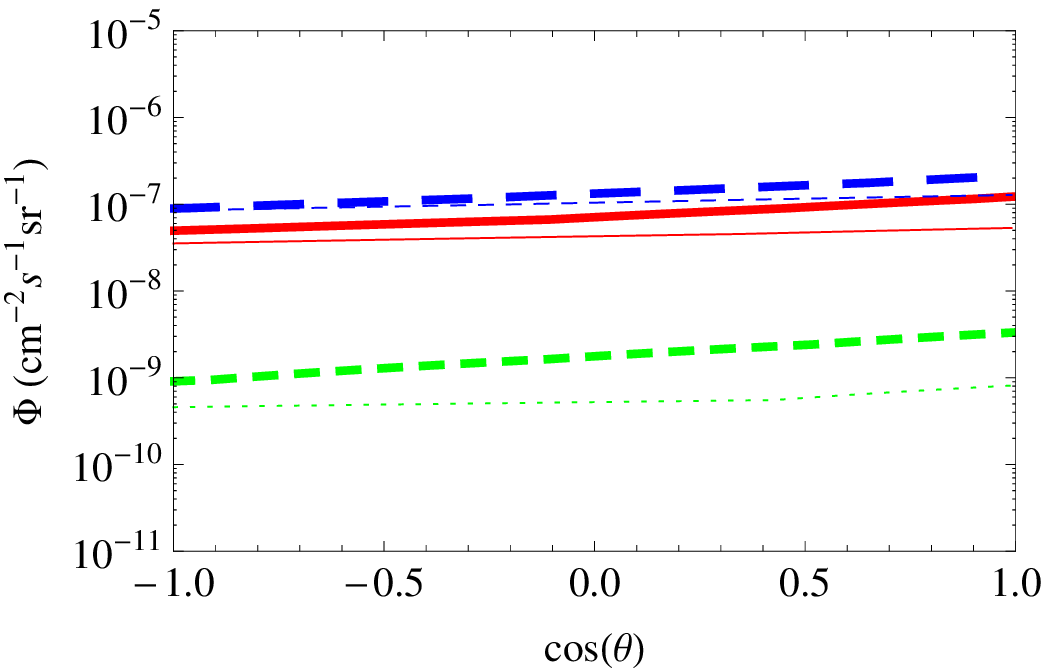,width=.4\textwidth}}\vspace{1mm}
\mbox{\epsfig{file=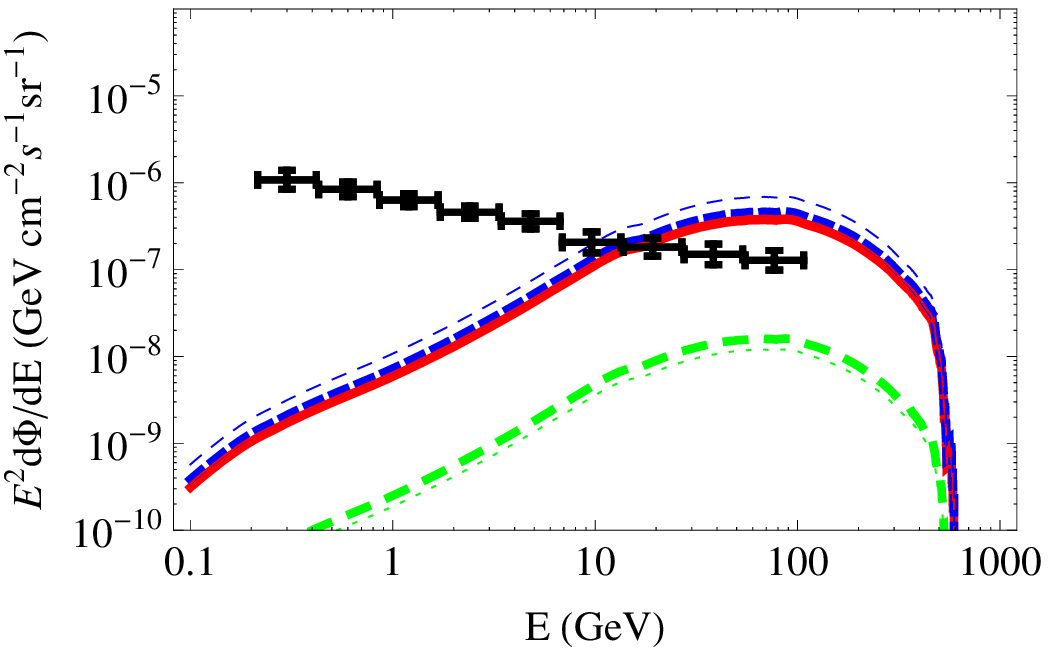,width=.4\textwidth}}\hspace{10mm}
\mbox{\epsfig{file=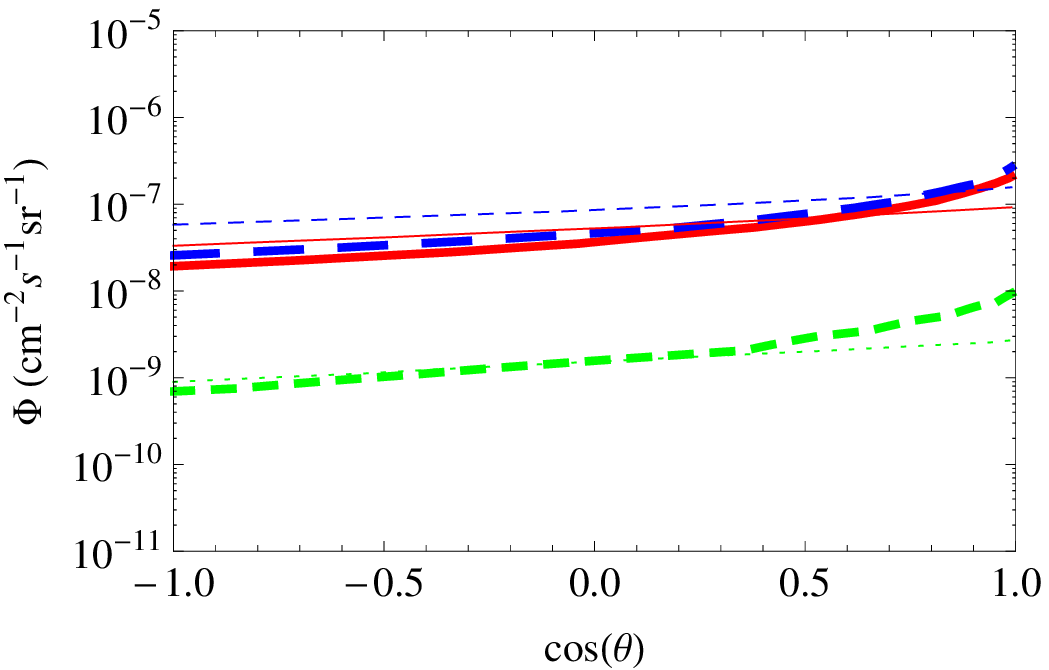,width=.4\textwidth}}\vspace{1mm}
\mbox{\epsfig{file=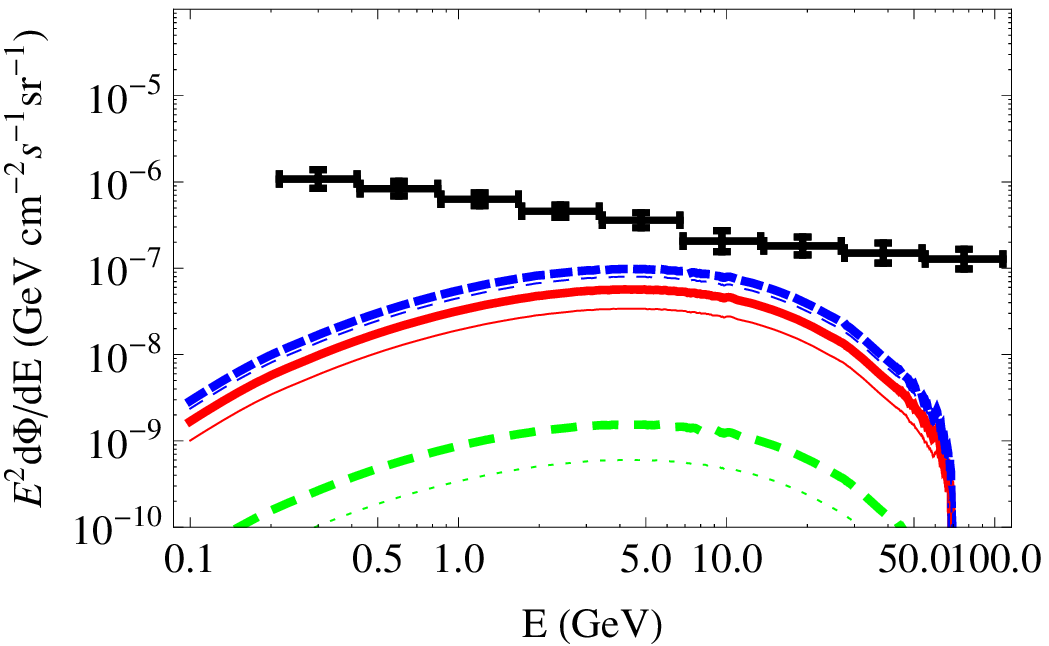,width=.4\textwidth}}\hspace{10mm}
\mbox{\epsfig{file=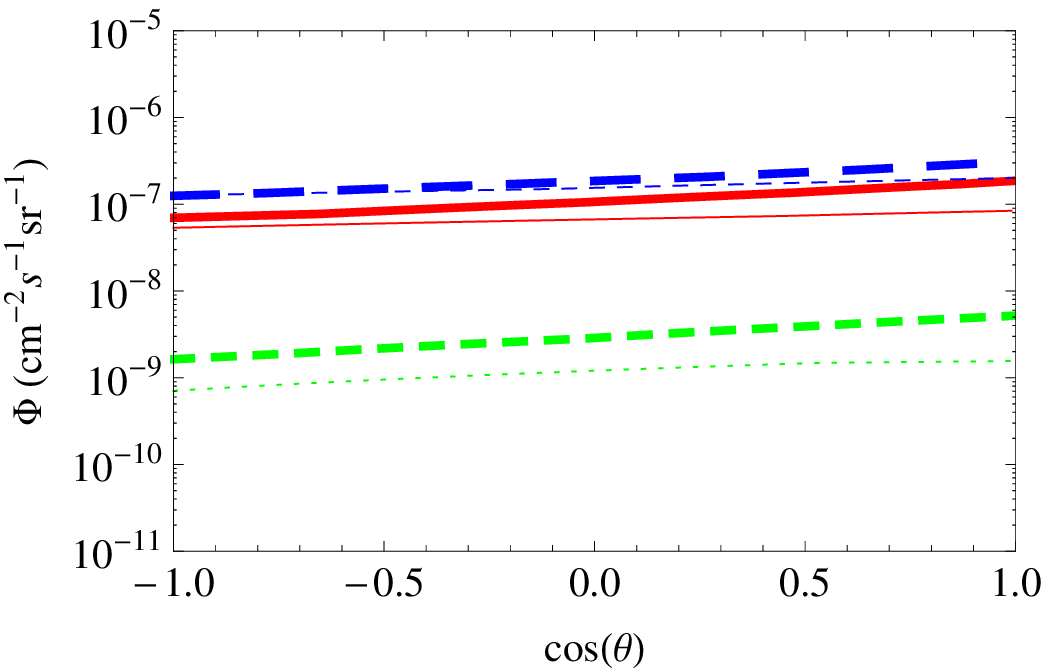,width=.4\textwidth}}\vspace{1mm}
\mbox{\epsfig{file=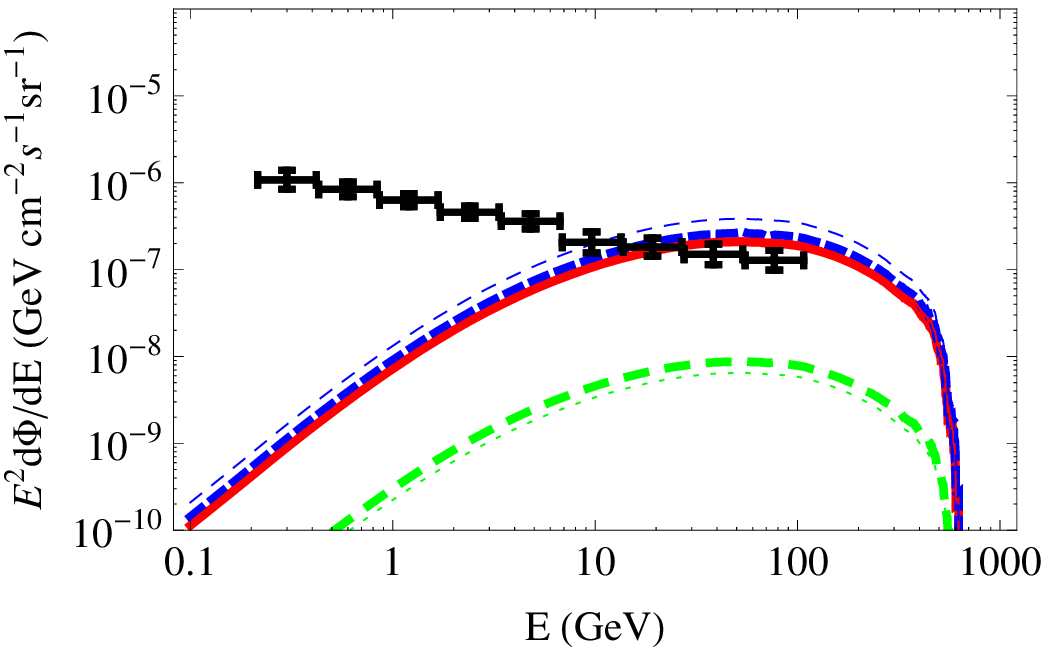,width=.4\textwidth}}\hspace{10mm}
\mbox{\epsfig{file=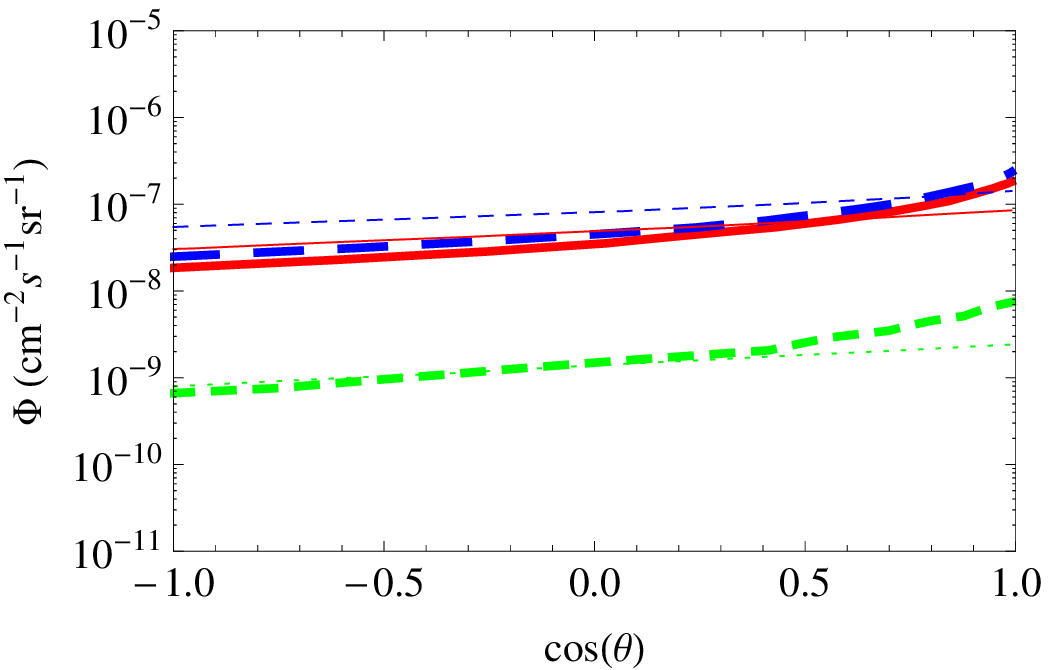,width=.4\textwidth}}
\end{center}
\caption{\it Diffuse gamma-ray flux from DM spikes for models b100, b1T, W100 and W1T, from top to bottom.  In each panel the Early (green, dotted), Intermediate (red, solid), and Late (blue, dashed) formation scenarios are shown for central black holes of mass $10^2 M_\odot$ (thick curves) and $10^3 M_\odot$ (thin curves).
In the left panels, the all-sky average flux is shown, with the FGST-measured EGB as black points~\cite{fgstEGB}.  In the right panels, the maximal anisotropy is displayed.  
\label{fig:Results1}}
\end{figure}

\begin{figure}[h!]
\begin{center}
\mbox{\epsfig{file=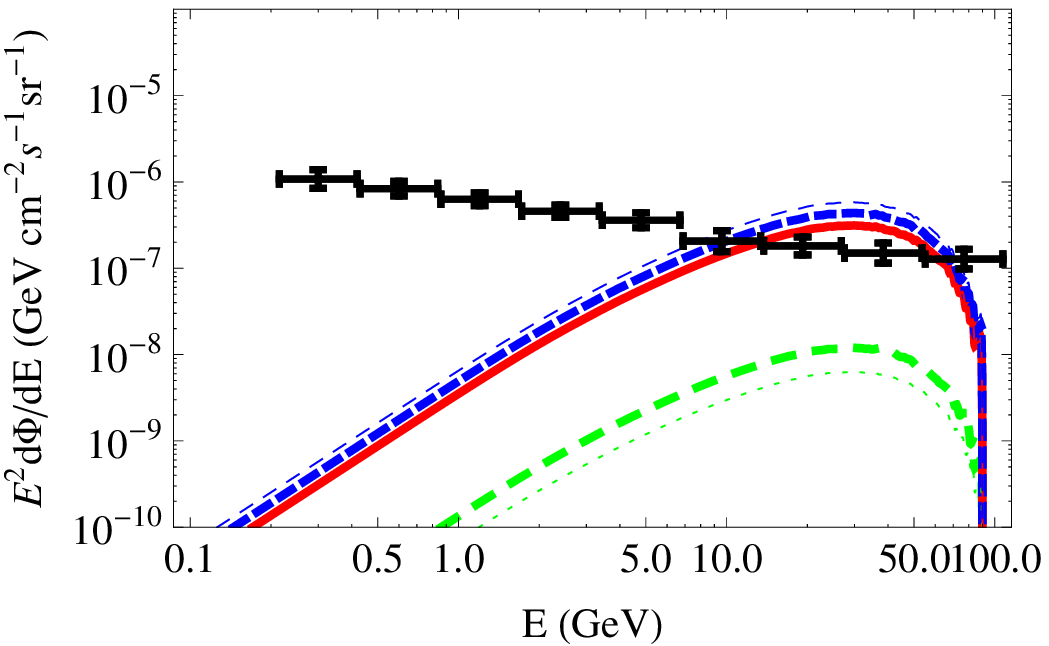,width=.4\textwidth}}\hspace{10mm}
\mbox{\epsfig{file=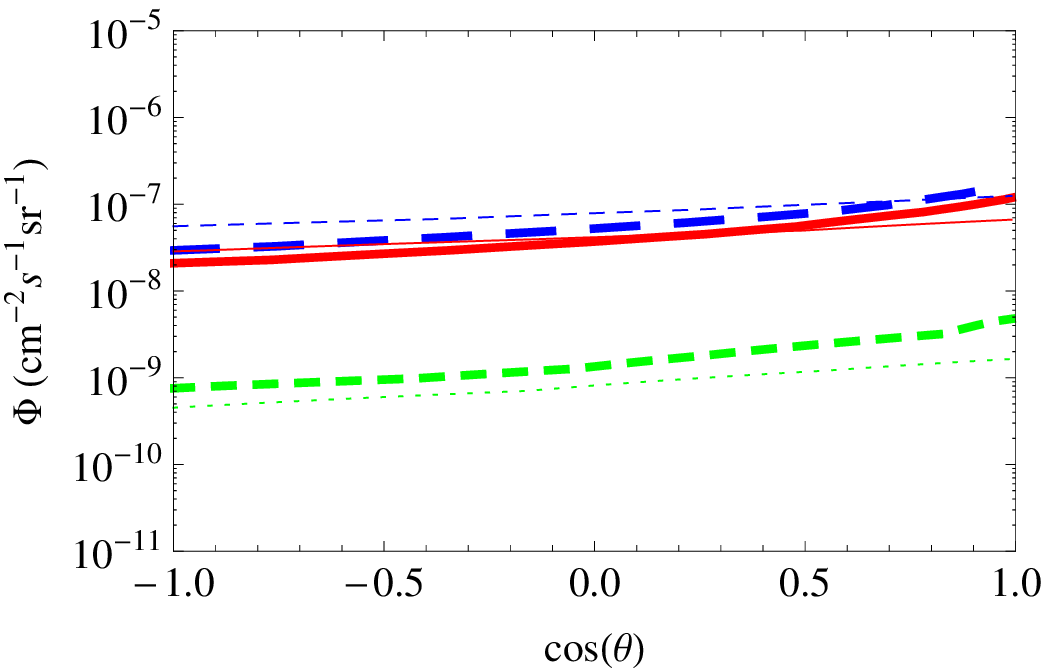,width=.4\textwidth}}\vspace{1mm}
\mbox{\epsfig{file=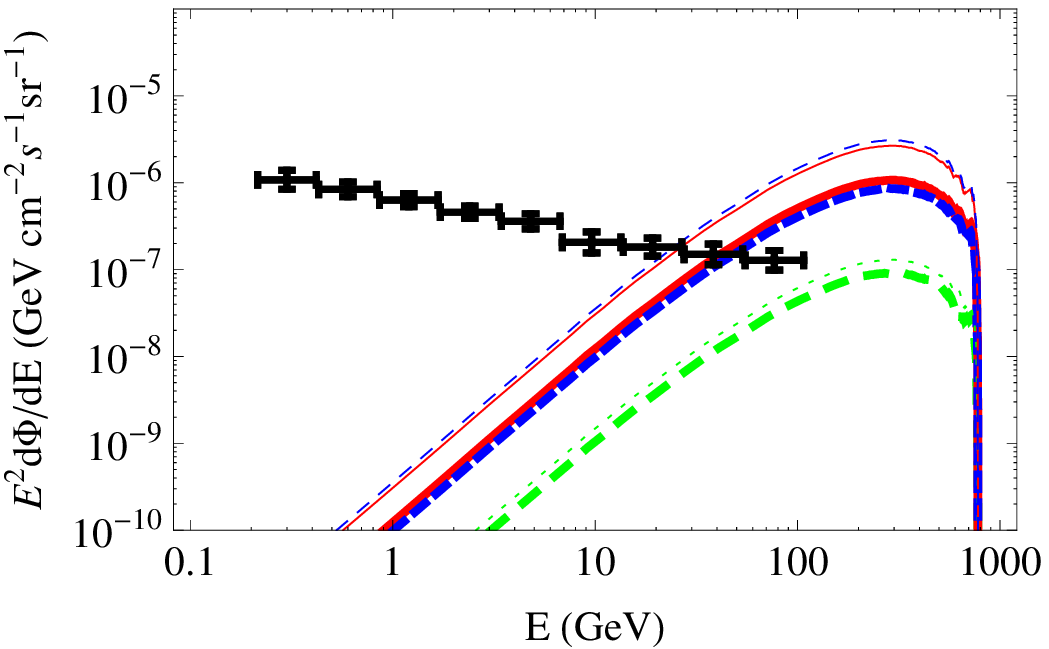,width=.4\textwidth}}\hspace{10mm}
\mbox{\epsfig{file=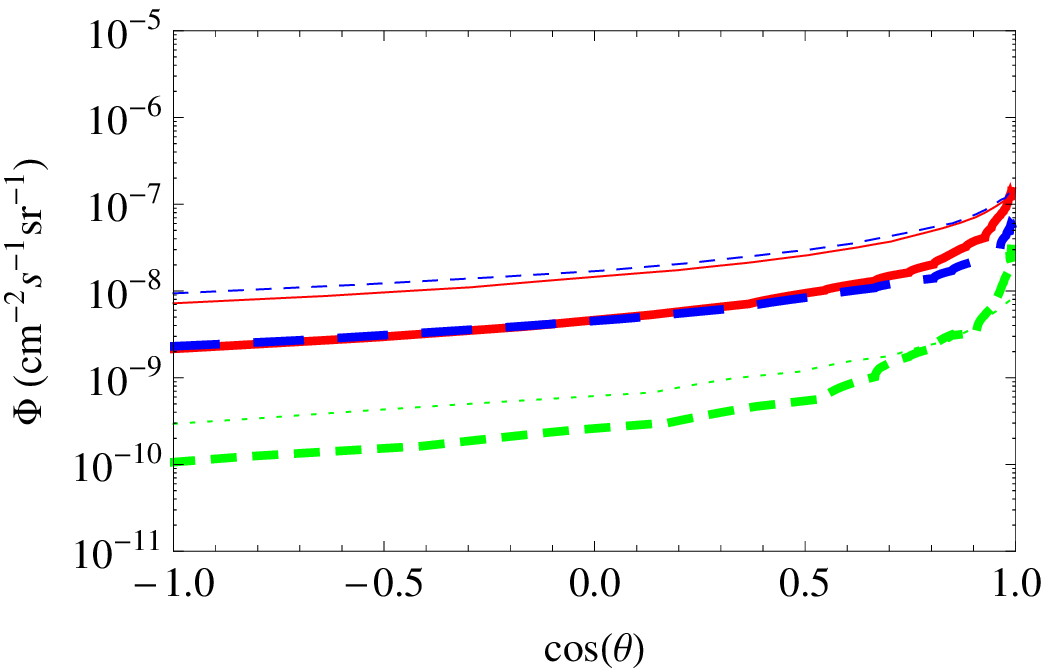,width=.4\textwidth}}\vspace{1mm}
\mbox{\epsfig{file=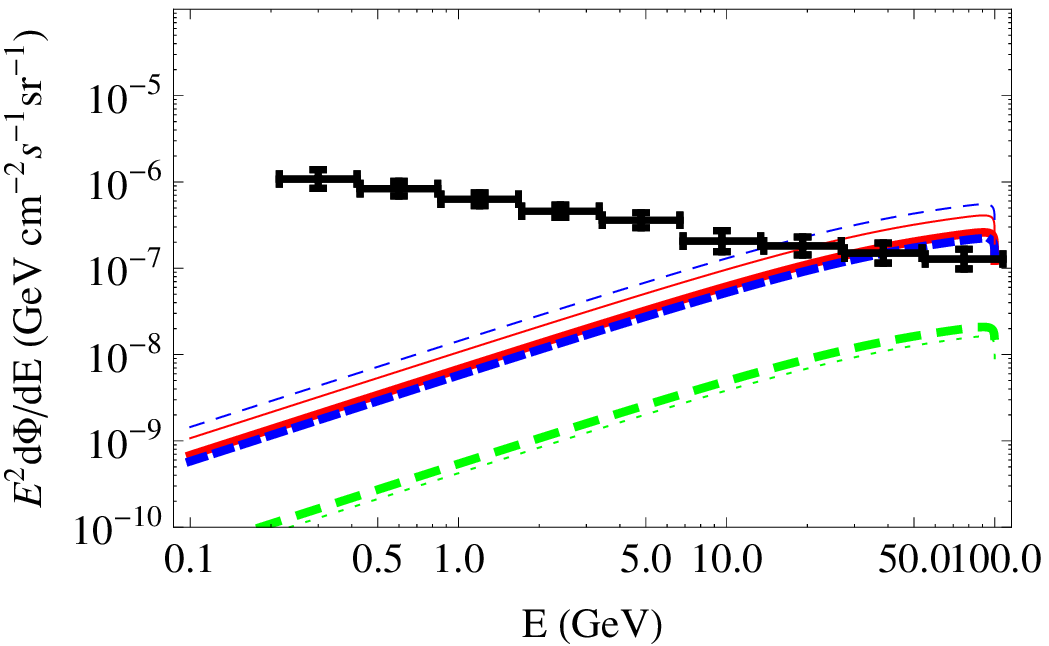,width=.4\textwidth}}\hspace{10mm}
\mbox{\epsfig{file=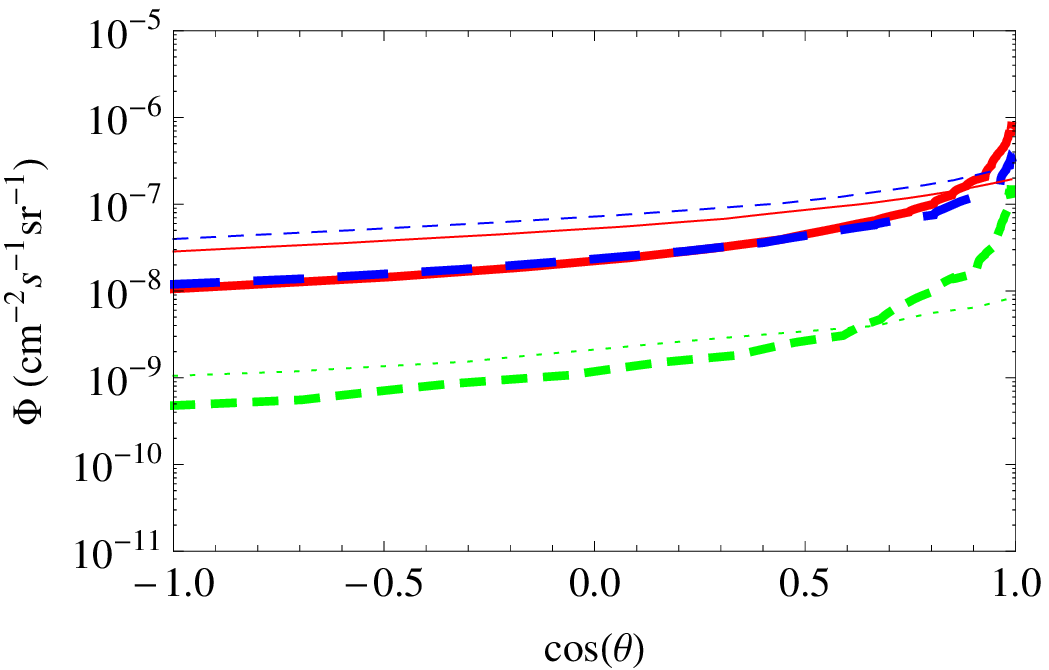,width=.4\textwidth}}\vspace{1mm}
\mbox{\epsfig{file=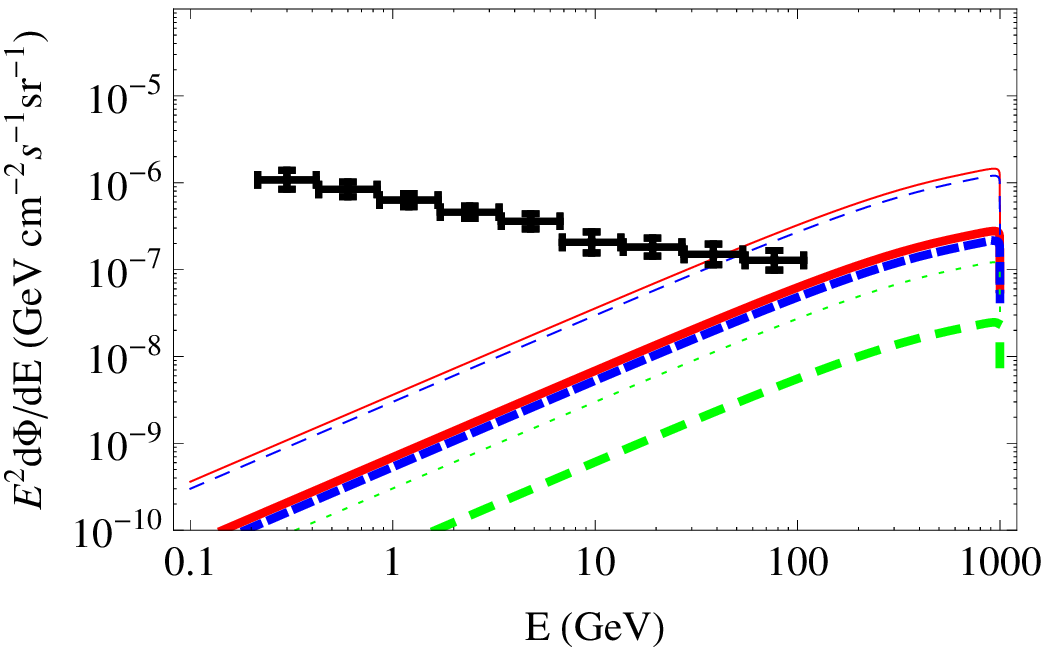,width=.4\textwidth}}\hspace{10mm}
\mbox{\epsfig{file=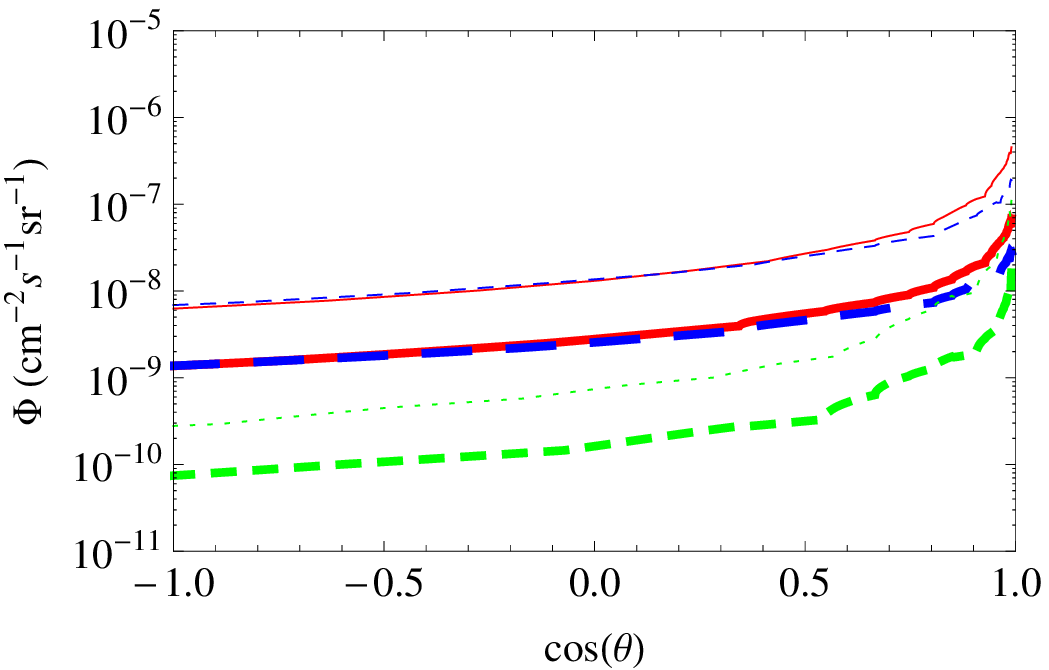,width=.4\textwidth}}
\end{center}
\caption{\it Diffuse gamma-ray flux from DM spikes for models $\tau$100, $\tau$1T, $\mu$100 and $\mu$1T, from top to bottom. Curves are as described in Fig.{\ref{fig:Results1}.}
\label{fig:Results2}}
\end{figure}

We see in the left panels of Figs.~\ref{fig:Results1} and~\ref{fig:Results2} that the all-sky averaged diffuse flux is well below the FGST measurement of the isotropic diffuse gamma-ray background in all models at low energies, but rises to the level of the FGST-measured flux at higher energies\footnote{For the sake of the clarity of Figs.~\ref{fig:Results1} and~\ref{fig:Results2} we have plotted only two choices of central black hole mass, $10^2$ and $10^3 M_\odot$. Larger central black holes may result in either larger or smaller gamma-ray fluxes, depending on the WIMP annihilation model, generally not differing by more than an order of magnitude from the $m_{BH}=10^2 M_\odot$ case.}.

The diffuse flux from our models is expected to be relatively low in the following two cases:
\hfil\break {\it Case I:}  If the flux from individual DM spikes is very low such that even nearby spikes are not bright enough to be identified as point sources, we find that the total flux from all spikes is then also well below the measured diffuse flux.  An
 example of a case which contributes little in the form of both point sources and the diffuse flux is the
  annihilation of heavy WIMPs in the DM spikes around very small black holes; Fig.~\ref{fig:IntLum} 
  illustrates the low luminosity of a single black hole in this example.  
\hfil\break {\it  Case II:} Alternatively, if a significant fraction of the spikes in the Milky Way halo are bright enough to have been discovered as point sources, then only spikes very far from our Solar System would be faint enough to escape identification as point sources and therefore contribute to the diffuse gamma-ray flux.  This is the case in many models where the individual point sources are very bright; when the black holes are very massive, when they formed very early, and if the dark matter is relatively light.  Examples of this include annihilations to $b\bar{b}$ or $W^+W^-$ around black holes more massive than $10^2 M_\odot$, as shown in the left panels of Fig.~\ref{fig:Results1}.

For scenarios that lie between the two extremes of {\it Case I} and {\it Case II}, the expected diffuse flux of gamma-rays from all unresolved DM spikes in the Milky Way halo may be significantly larger than the FGST-measured diffuse flux.  Therefore, for each black hole mass, $z_f$, and dark matter mass and annihilation mode,   
a constraint on $f_{DS}$ may be derived from the requirement that the diffuse flux of gamma-rays from DM spikes not produce a significant excess over the FGST observation.  These constraints will be discussed in Section~\ref{sec:fDS}.

It is also interesting to examine the anisotropy of the expected 
diffuse gamma-ray flux, namely to study the number of events as a 
function of angle with respect to the Solar System-Galactic center 
axis.  This anisotropy is plotted in the right panels of 
Figs.~\ref{fig:Results1} and~\ref{fig:Results2}.   Previously it was 
suggested that the anisotropy might be used to discriminate among 
models of Milky Way substructure and dark matter 
annihilations~\cite{aloisio}.  Though such an analysis has not yet 
been done with data from FGST, here we consider the behavior of the 
anisotropy as an indication of whether DM spikes near the Galactic 
center contribute to the diffuse gamma-ray flux.

For very low luminosity spikes, as in {\it Case I}, few, if any, 
spikes are bright enough to have been identified as point sources.  If 
this is the case, then nearly all spikes contribute to the diffuse 
signal, including those near the Galactic center.  The diffuse 
gamma-ray flux would therefore be peaked towards to Galactic center.   
As one can see in the right panels of Figs.~\ref{fig:Results1} 
and~\ref{fig:Results2}, for most dark matter models shown, the spikes 
surrounding $10^2 M_\odot$ black holes (thick curves) are indeed so 
dim that those from the Galactic center region contribute to the 
diffuse flux.  The exceptions to this are the two most luminous dark 
matter models shown, $b100$ and $W100$, which exhibit more isotropic 
signals.

All other factors being equal, the DM spikes surrounding $10^3 
M_\odot$ black holes are nearly an order of magnitude brighter than 
those surrounding $10^2 M_\odot$ black holes.  These higher-luminosity 
DM spikes are generally bright enough that the closest ones must be at 
least as far from our Solar System as the Galactic center, as 
evidenced by the lack of peaking toward the Galactic center for the 
thin curves in the right panels of Figs.~\ref{fig:Results1} 
and~\ref{fig:Results2}.  The exceptions that exhibit a peak towards 
the Galactic center for $10^3 M_\odot$ black holes are models $\mu1T$ 
and $\tau1T$, the least luminous models considered here. For even 
larger black holes, not shown in Figs.~\ref{fig:Results1} 
and~\ref{fig:Results2}, the luminosity of each spike may be so high 
that only spikes in the very outer regions of the Milky Way halo (if 
any) contribute to the diffuse gamma-ray flux, as in {\it Case II}. 
This results in a nearly-completely isotropic diffuse flux.

In contrast to the magnitude of the diffuse flux, which would be small 
in both {\it Case I} and {\it Case II}, the anisotropy is expected to 
be different in the two cases: {\it Case I} would result in some 
peaking of the signal towards the Galactic center, while the signal 
from {\it Case II} would be nearly isotropic.  However, it must be 
pointed out that any anisotropy would be most evident near the 
Galactic center, where there are many baryonic gamma-ray sources, 
making the diffuse component difficult to disentangle, and where the 
survival of DM spikes is less certain. If spikes in the central region 
were disrupted, any anisotropy may have been washed out. Therefore, if 
there is a peak in the angular distribution of the diffuse gamma-ray 
flux towards the Galactic center, it may be evidence of a {\it Case I}-like 
population of DM spikes, but the lack of a strong anisotropy is not 
necessarily evidence for a {\it Case II}-like population.

In all scenarios discussed above, the star formation history plays a significant role in so far as the total number of DM spikes in the Milky Way halo is greatest if star formation persisted to low redshift.  If Population III.1 star formation ended at high redshift, as for Early $z_f$, there would be significantly fewer black hole remnants today in the Milky Way halo.  This scenario, identified by the green curves in Figs.~\ref{fig:Results1} and~\ref{fig:Results2}, universally results in a lower diffuse flux than the other scenarios examined here.  
The Intermediate and Late $z_f$ scenarios, identified by the red and blue curves, respectively, 
typically have results that are quite similar to each other, both for the total gamma-ray fluxes and for the spectra.
Again, we emphasize that we have taken $f_{DS}=1$ in these calculations.

\section{Constraining $f_{DS}$}
\label{sec:fDS}

\begin{figure}[h!]
\begin{center}
\mbox{\epsfig{file=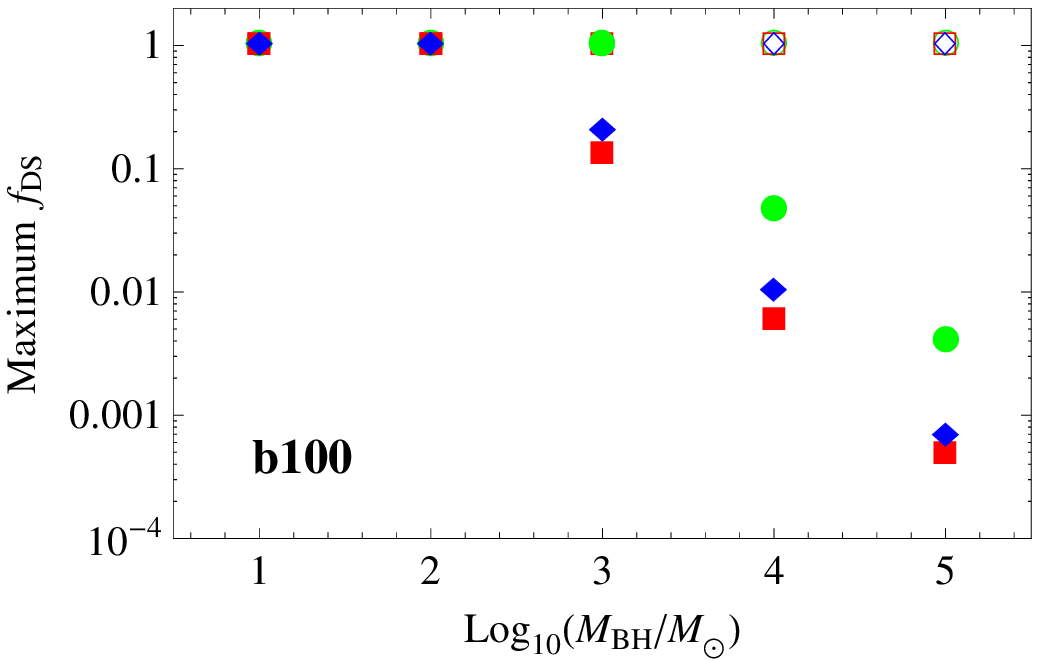,width=.4\textwidth}}\hspace{10mm}
\mbox{\epsfig{file=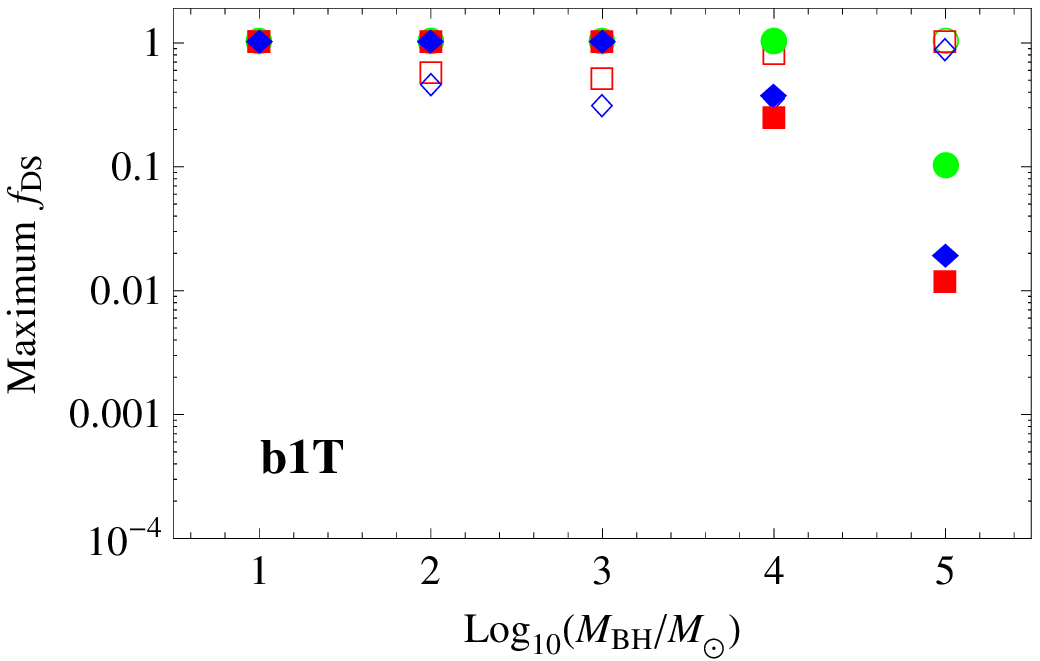,width=.4\textwidth}}\vspace{2mm}
\mbox{\epsfig{file=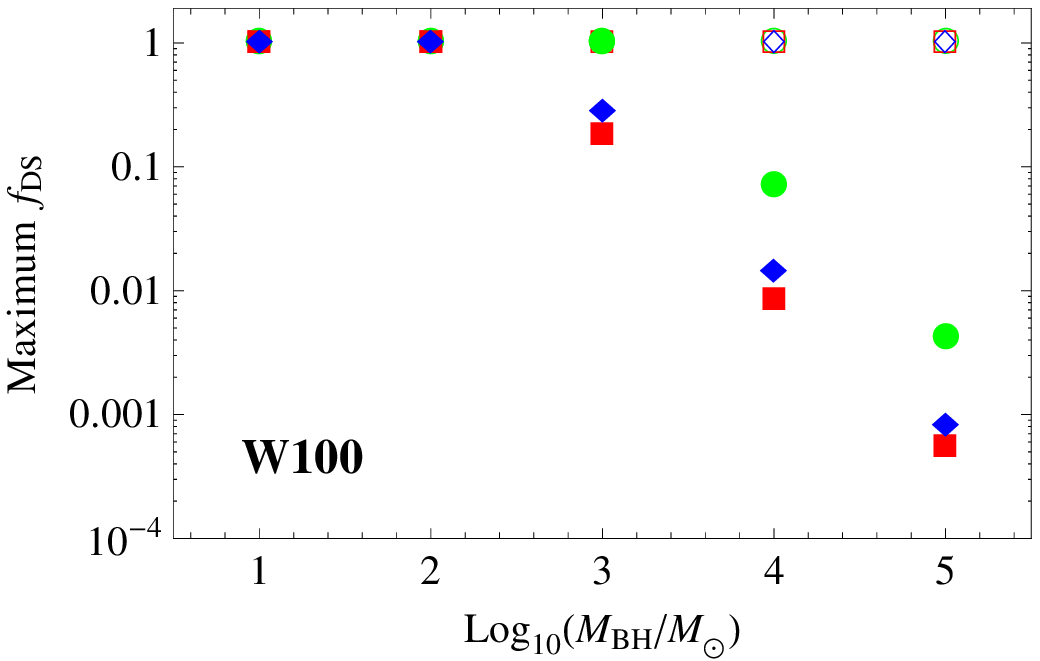,width=.4\textwidth}}\hspace{10mm}
\mbox{\epsfig{file=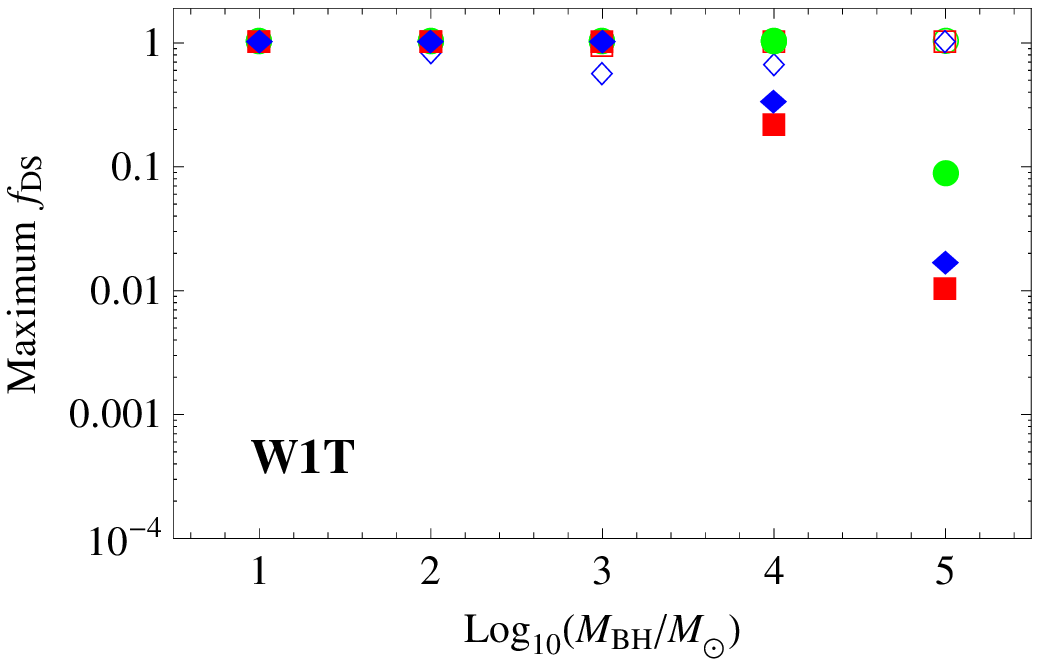,width=.4\textwidth}}\vspace{2mm}
\mbox{\epsfig{file=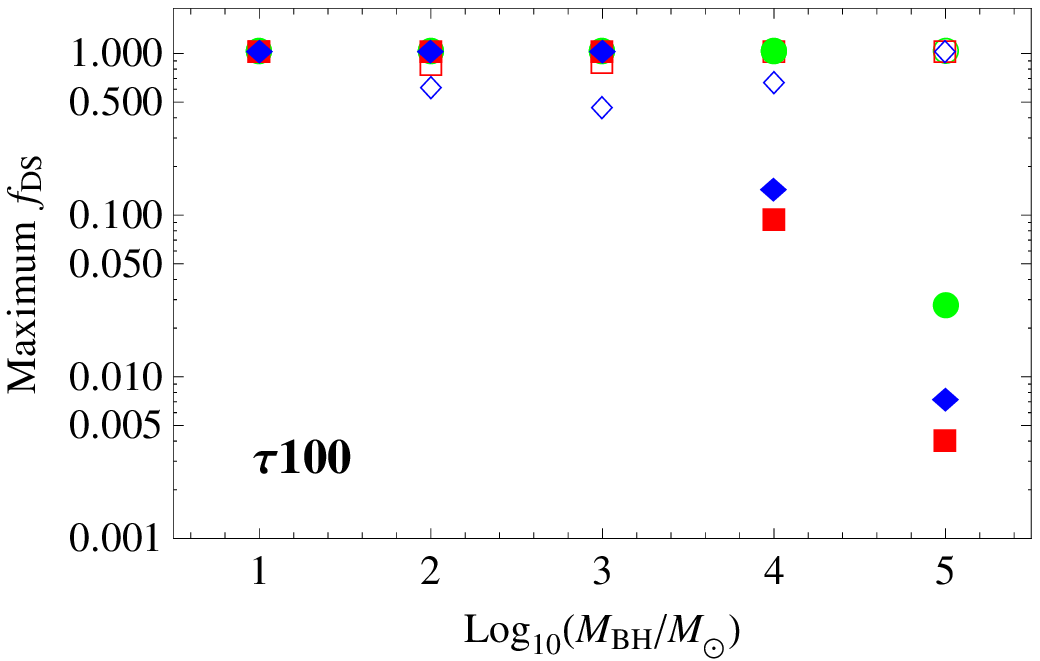,width=.4\textwidth}}\hspace{10mm}
\mbox{\epsfig{file=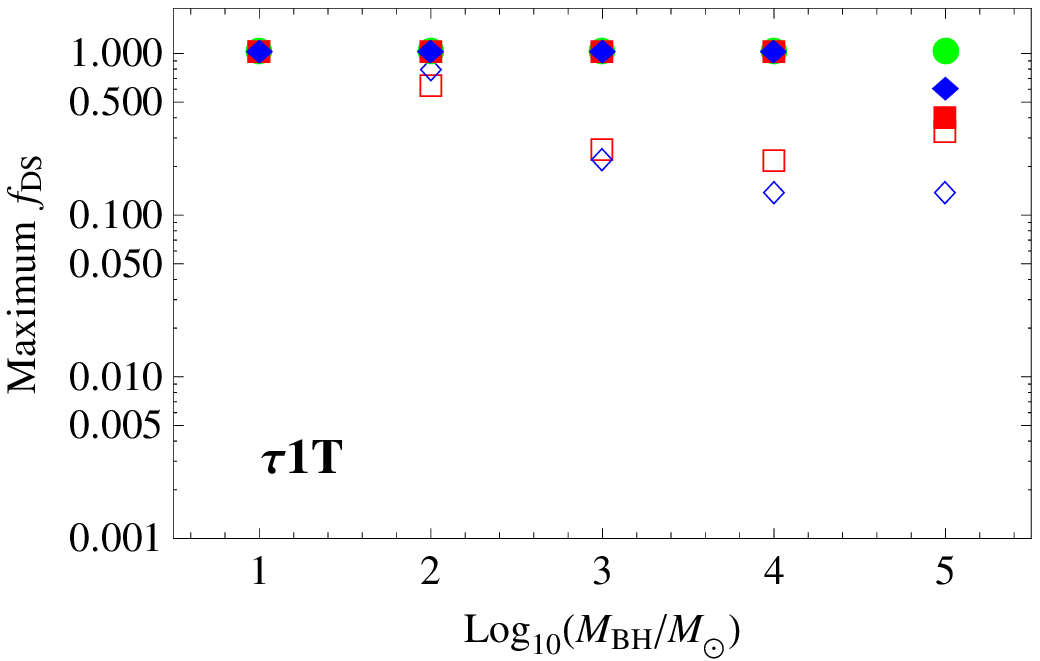,width=.4\textwidth}}\vspace{2mm}
\mbox{\epsfig{file=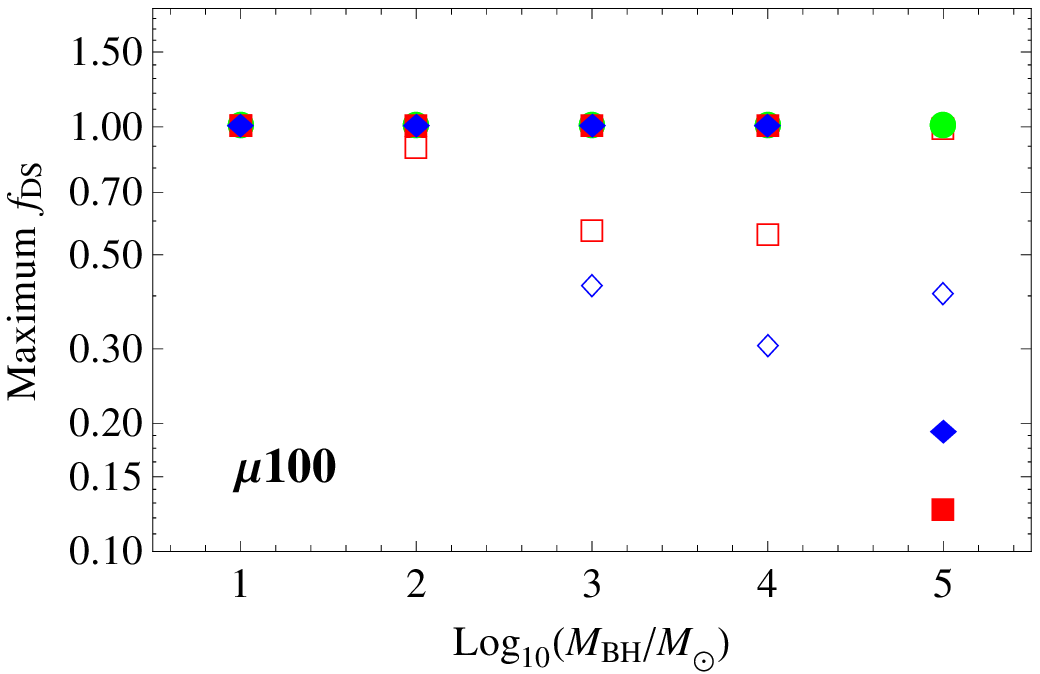,width=.4\textwidth}}\hspace{10mm}
\mbox{\epsfig{file=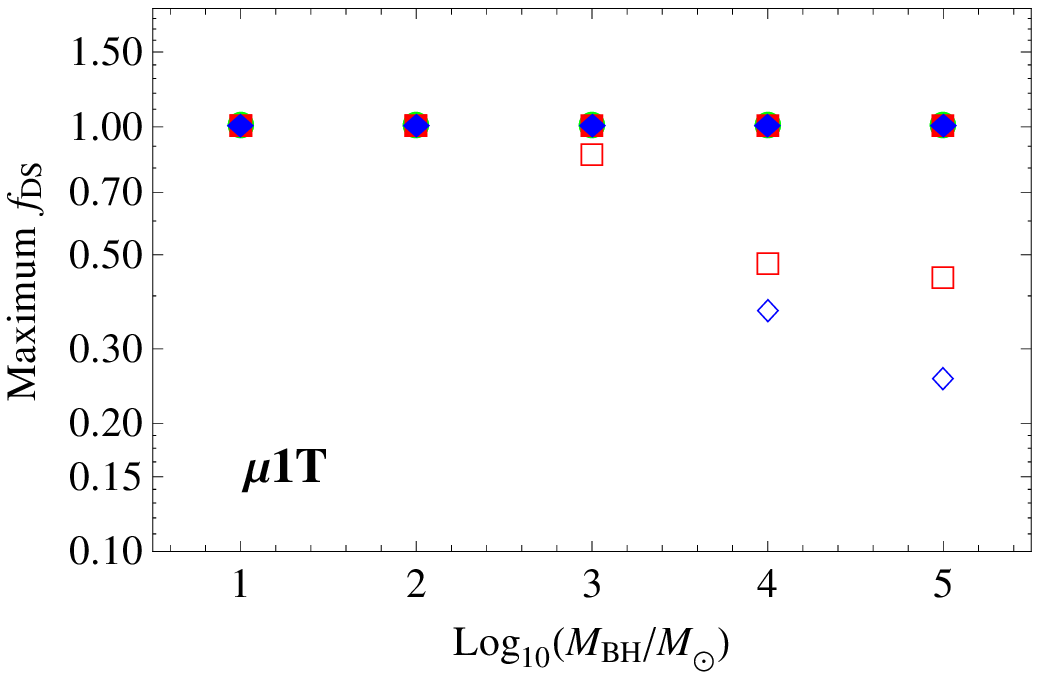,width=.4\textwidth}}
\end{center}
\caption{\it Maximum $f_{DS}$ as a function of central black hole mass for the WIMP annihilation models in Table~\ref{tab:wimpmodels}. Green circles, red squares, and blue diamonds are for Early, Intermediate, and Late $z_f$, respectively. The solid markers are the limits from point source brightness, while the open markers are from the diffuse gamma-ray flux, as described in the text.  In each panel, the branching fraction to the relevant final state
$B_f=1$. Note that the range of $f_{DS}$ displayed differs from panel to panel.
\label{fig:fDSmax}}
\end{figure}

To this point we have focused mainly on $f_{DS}=1$, though it is possible that $f_{DS} \ll 1$.  It is possible to constrain $f_{DS}$ in two ways:  First, from the FGST measurement of the diffuse gamma-ray background, we require that the diffuse flux from dark matter annihilations around spikes in the Milky Way halo not exceed the measured flux in any of the nine FGST energy bins in Ref.~\cite{fgstEGB} by more than $3\sigma$.  The diffuse gamma-ray flux from annihilations to final state $f$ may be rewritten for $f_{DS}\neq 1$ as 
\beq
\Phi_f (f_{DS}) = f_{DS} \times \Phi_f(f_{DS}=1).
\eeq
One can then extract an upper limit on $f_{DS}$ from the diffuse background for each choice of $z_f$ and central black hole mass in each of the dark matter annihilation models.  The resulting maximal values of $f_{DS}$ are presented in Fig.~\ref{fig:fDSmax} for each of the WIMP annihilation models in Table~\ref{tab:wimpmodels} as the open points, with Early, Intermediate, and Late $z_f$ represented in green, red, and blue, respectively.   Note that some scenarios are entirely unconstrained (maximal $f_{DS} = 1$), in which case the markers are superposed on each other.
These constraints are strongest for scenarios in which the intrinsic luminosity of individual DM spikes is moderate; that is, bright enough that the signal is strong, but not so bright that many/all spikes in the Milky Way halo are seen as point sources.  The constraints are also weakest in all cases for Early termination of star formation, as there are fewer potential nearby DM spikes today.

The second way to constrain $f_{DS}$ is by requiring that one should expect to find less than one DM spike within $D_{min}^{PS}$, the minimal distance at which a spike may be located such that it is not brighter than the brightest point source in the FGST First Source Catalog. 
In this case we rewrite the number density of DM spikes as a function of distance from the Galactic center, $N_{sp}(R)$, as
\beq
N_{sp}(R, f_{DS}) = f_{DS} \times N_{sp}(R, f_{DS}=1),
\eeq
and find $f_{DS}$ such that
\beq
\int_0^{D_{min}^{PS}} ds \int_{\text{all sky}} d\Omega \, N_{sp}(R,f_{DS}) \leq 1.
\label{eq:PSconstraint}
\eeq
The maximum $f_{DS}$ as found in this method is displayed in Fig.~\ref{fig:fDSmax} as the solid points in each scenario.  These are the most conservative limits on $f_{DS}$, as they are based on the flux not exceeding that of the brightest FGST point source, the Vela pulsar.  If we require that the flux from a DM spike not exceed that of the brightest unassociated point source, stronger limits on $f_{DS}$ would be obtained, though there is no gaurantee that there is not a DM spike along our line of sight to Vela. 

We see from the solid points in Fig.~\ref{fig:fDSmax} that the point source constraint on $f_{DS}$ is strongest in all cases for large central black hole masses and low WIMP masses, when the intrinsic luminosity of each individual spike is largest.  Again, they are weak in the case of Early termination of star formation, as there are fewer black holes expected in the Milky Way halo, and therefore fewer in the vicinity of our Solar System. 

\begin{figure}[h!]
\begin{center}
\mbox{\epsfig{file=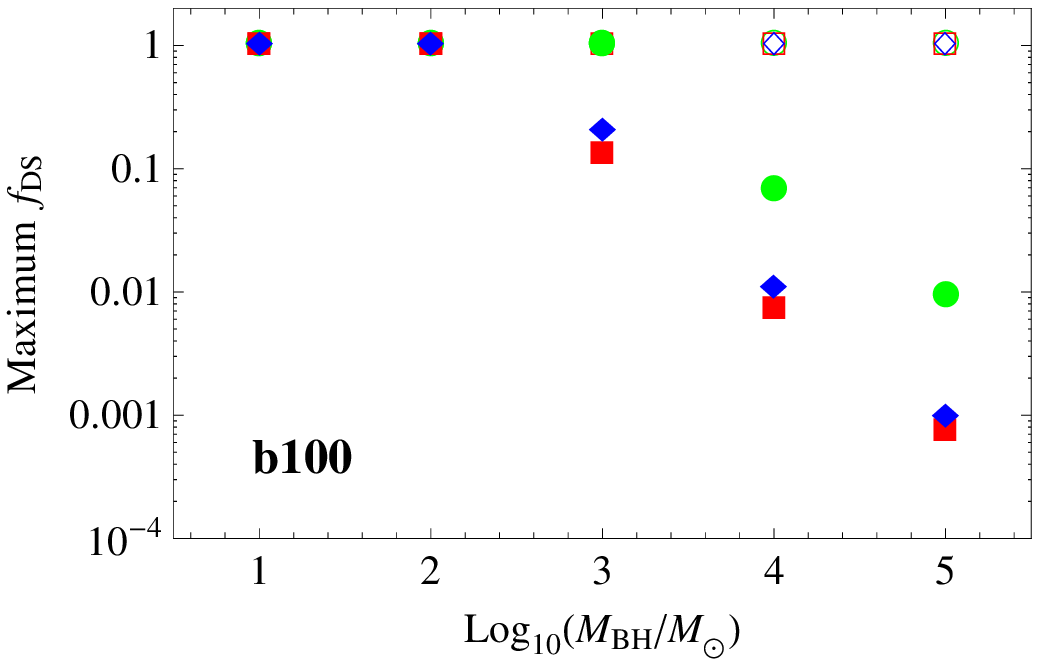,width=.4\textwidth}}\hspace{10mm}
\mbox{\epsfig{file=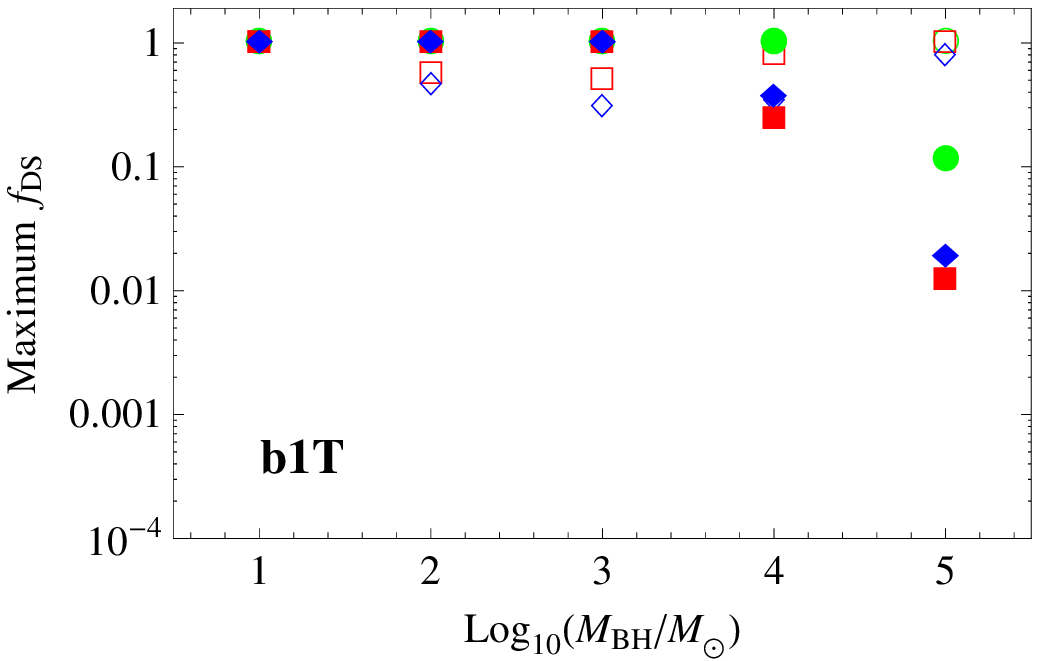,width=.4\textwidth}}\vspace{2mm}
\mbox{\epsfig{file=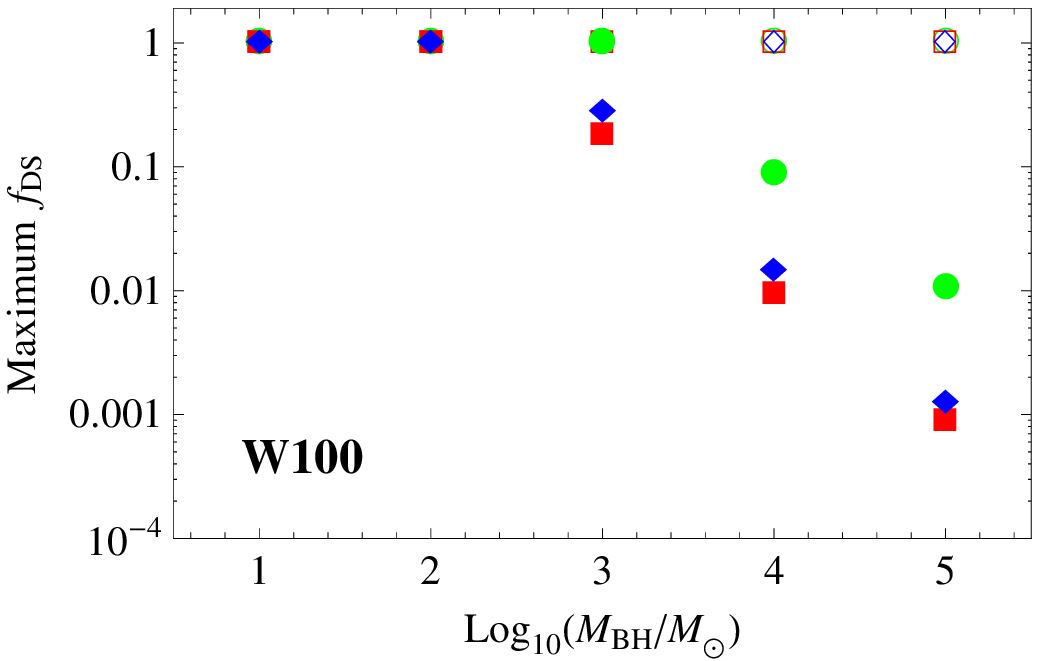,width=.4\textwidth}}\hspace{10mm}
\mbox{\epsfig{file=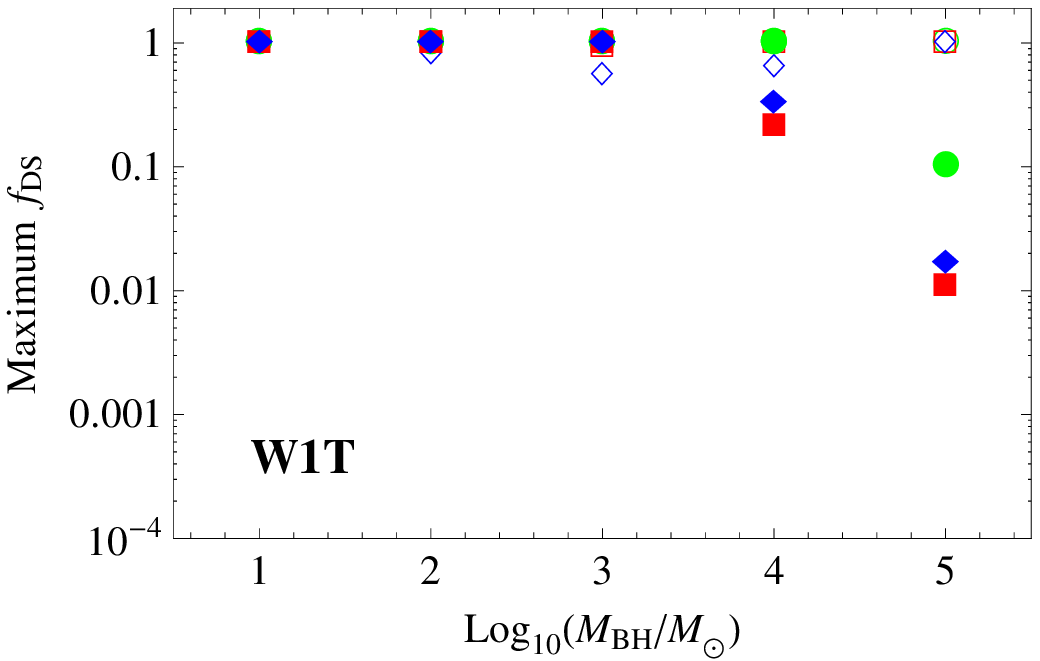,width=.4\textwidth}}\vspace{2mm}
\mbox{\epsfig{file=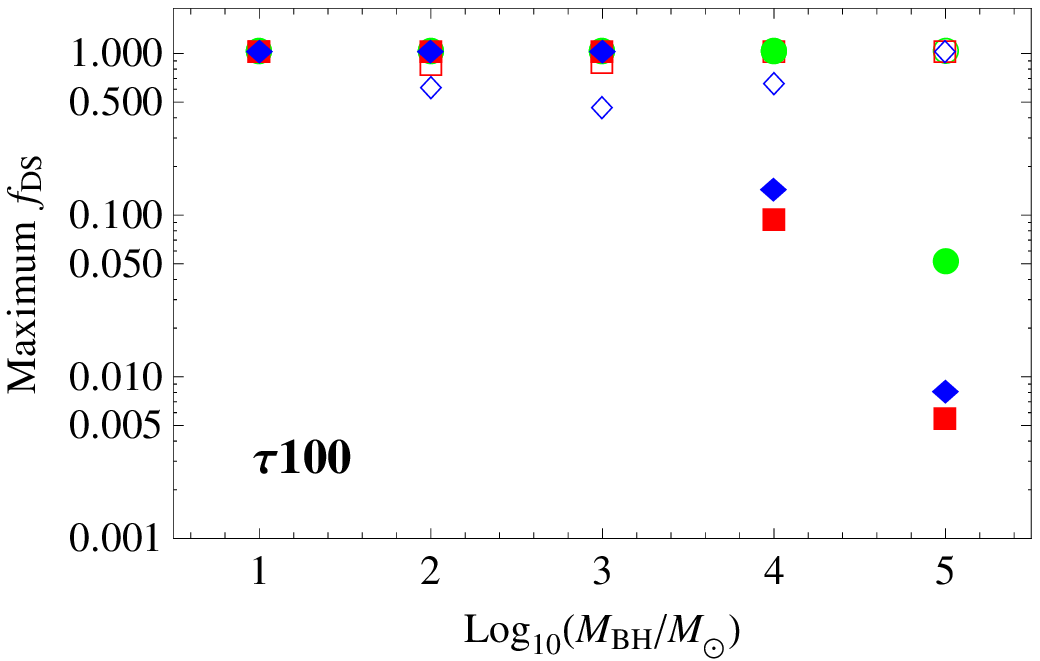,width=.4\textwidth}}\hspace{10mm}
\mbox{\epsfig{file=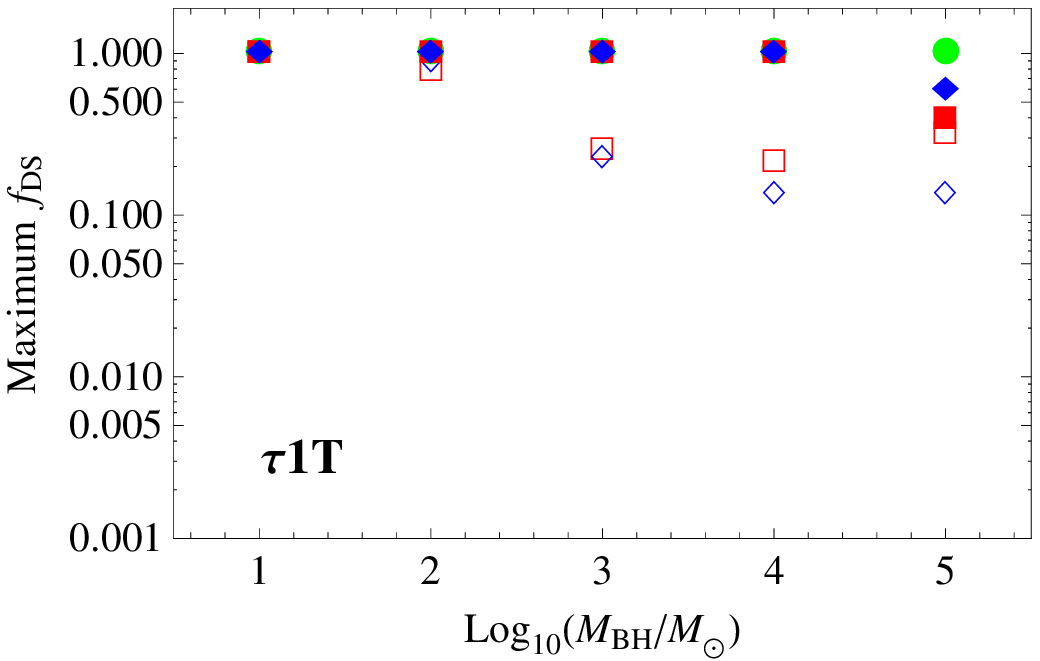,width=.4\textwidth}}\vspace{2mm}
\mbox{\epsfig{file=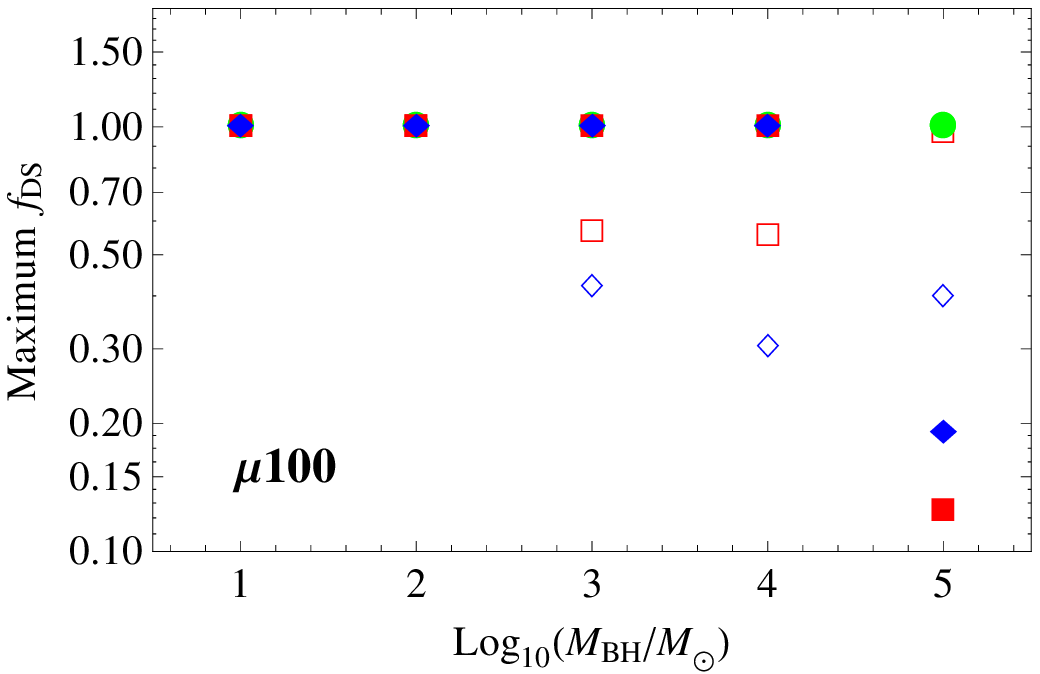,width=.4\textwidth}}\hspace{10mm}
\mbox{\epsfig{file=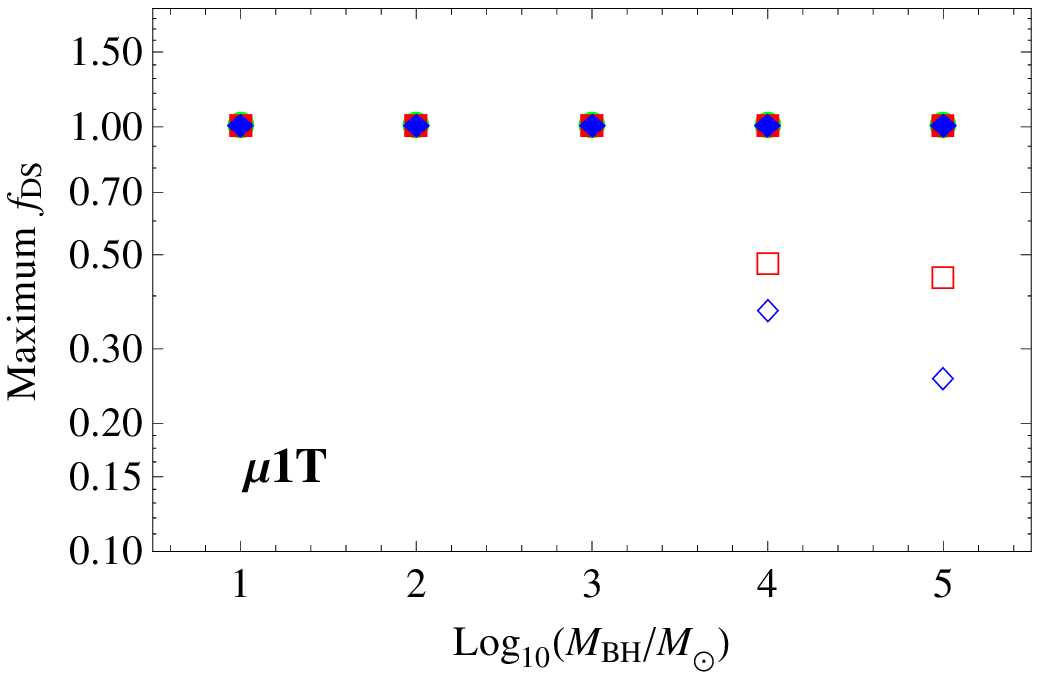,width=.4\textwidth}}
\end{center}
\caption{\it Maximum $f_{DS}$ as in Fig.~\ref{fig:fDSmax}, but excluding the inner 5 kpc from the Galactic center.
\label{fig:fDSmax5}}
\end{figure}

To this point we have not properly addressed the uncertainty in the DM spike distribution near the Galactic center.  In fact, in the central region of the Milky Way halo, subhalo evolution is affected by processes such as dynamical friction and tidal mass loss~\cite{zentner2007}, resulting in a dearth of dark matter substructures at small Galactic radii.  However, as pointed out in Ref.~\cite{bzs}, black holes and their surrounding dark matter spikes can survive tidal disruption even near the Galactic center.  Still, Ref.~\cite{bzs} finds that the distribution of DM spikes is very uncertain, and potentially even zero, in the inner $\sim 3$ kpc (see, for example, Fig.~1 of~\cite{bftz}). 

Given this uncertainty, we calculate also the constraints on $f_{DS}$ in each scenario, neglecting all DM spikes in the inner 5 kpc of the Galaxy. In Fig.~\ref{fig:fDSmax5}, these results are presented.  If $f_{DS}=1$, for Early, Intermediate, and Late $z_f$, we expect roughly 154, 407, and 895 DM spikes in the inner 5 kpc, corresponding to 38\%, 10\%, and 4\% of the total number of DM spikes in the Milky Way halo, respectively. The constraints are strikingly similar to those in Fig.~\ref{fig:fDSmax}.  In general, the limits on $f_{DS}$ from point source brightness come from much smaller distances, as can be seen in Fig.~\ref{fig:PointSourceDist}, while the limits from the diffuse flux are sensitive primarily to the bulk of the halo at large Galactic radii.  As a result, there are only a few cases in which the constraints visibly differ, all of which concern the nearest point source in scenarios where individual DM spikes are bright enough that $D_{min}^{PS} \gtrsim 8.5$ kpc.  With the exception  of these few cases, our constraints on $f_{DS}$ are robust with respect to the distribution of DM spikes near the Galactic center.

\begin{figure}[h!]
\begin{center}
\mbox{\epsfig{file=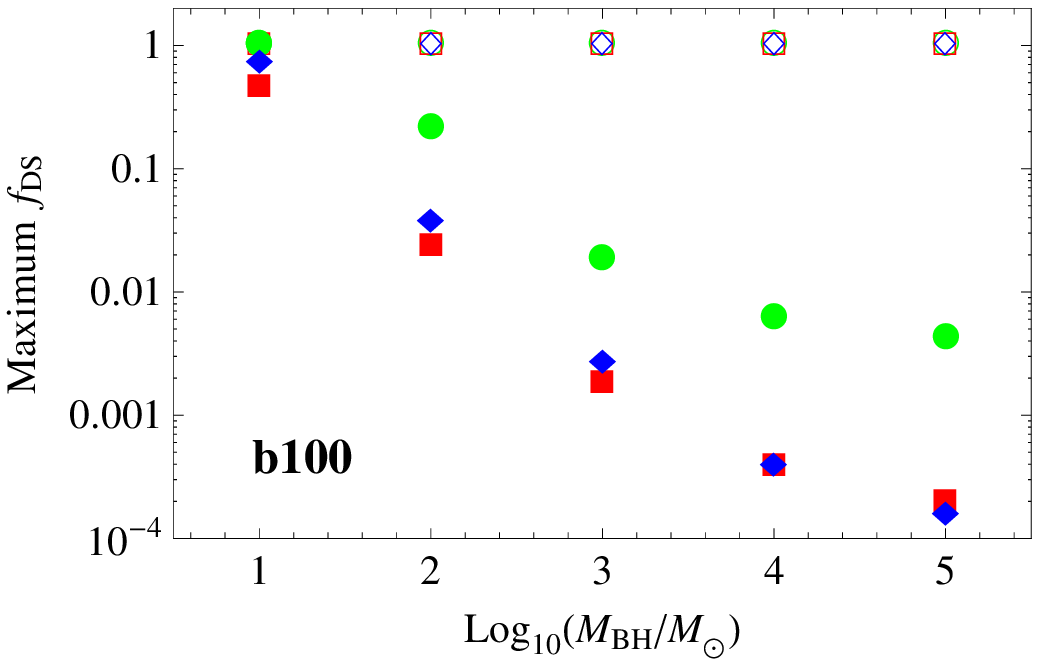,width=.4\textwidth}}\hspace{10mm}
\mbox{\epsfig{file=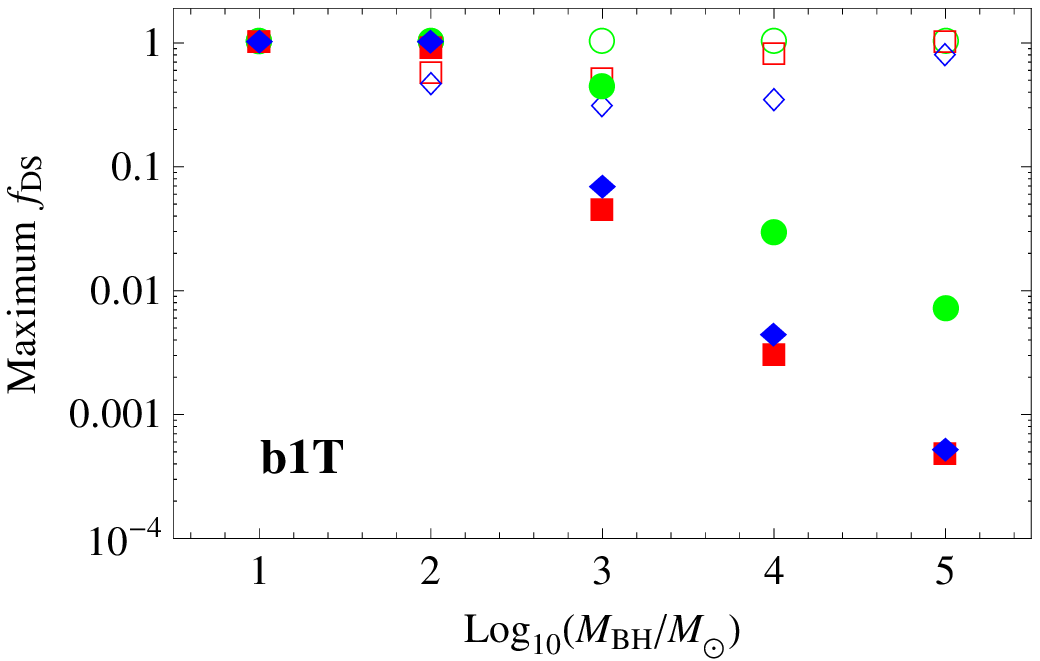,width=.4\textwidth}}\vspace{2mm}
\mbox{\epsfig{file=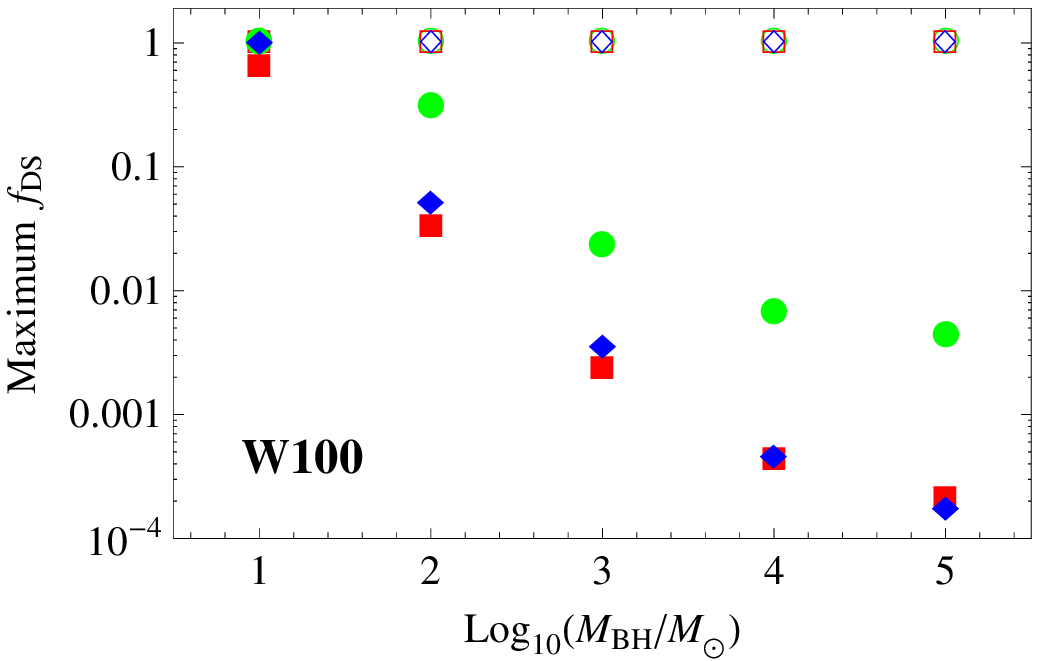,width=.4\textwidth}}\hspace{10mm}
\mbox{\epsfig{file=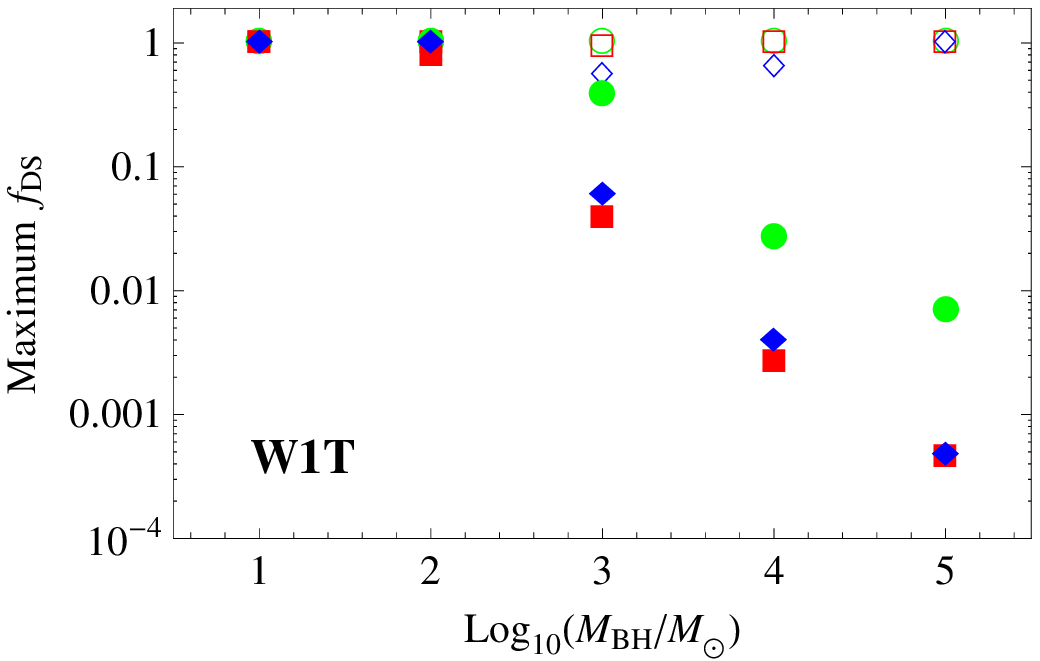,width=.4\textwidth}}\vspace{2mm}
\mbox{\epsfig{file=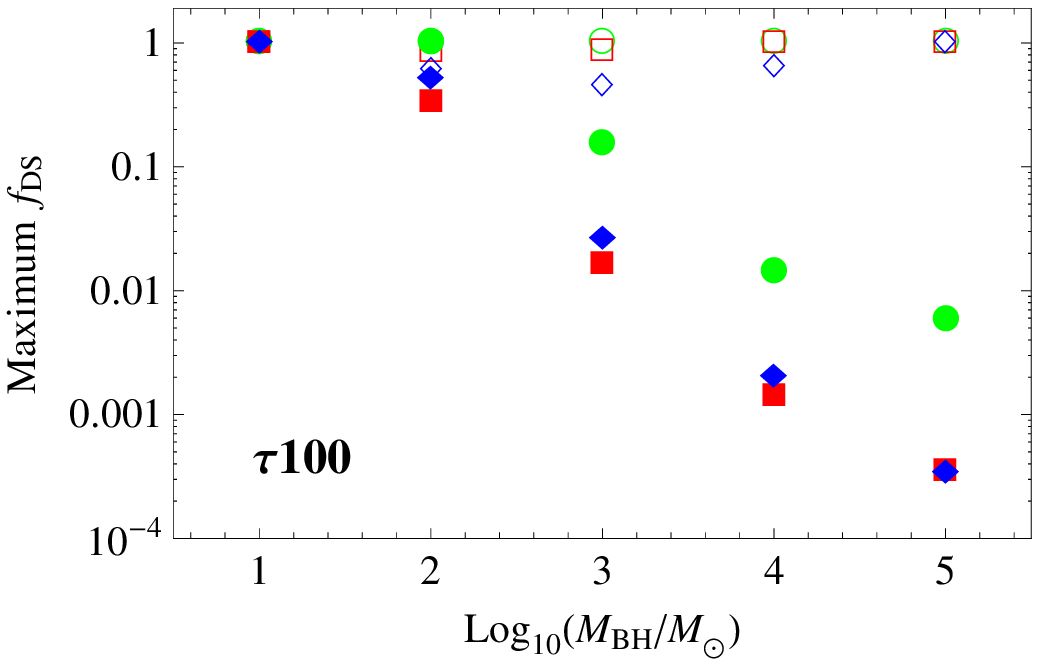,width=.4\textwidth}}\hspace{10mm}
\mbox{\epsfig{file=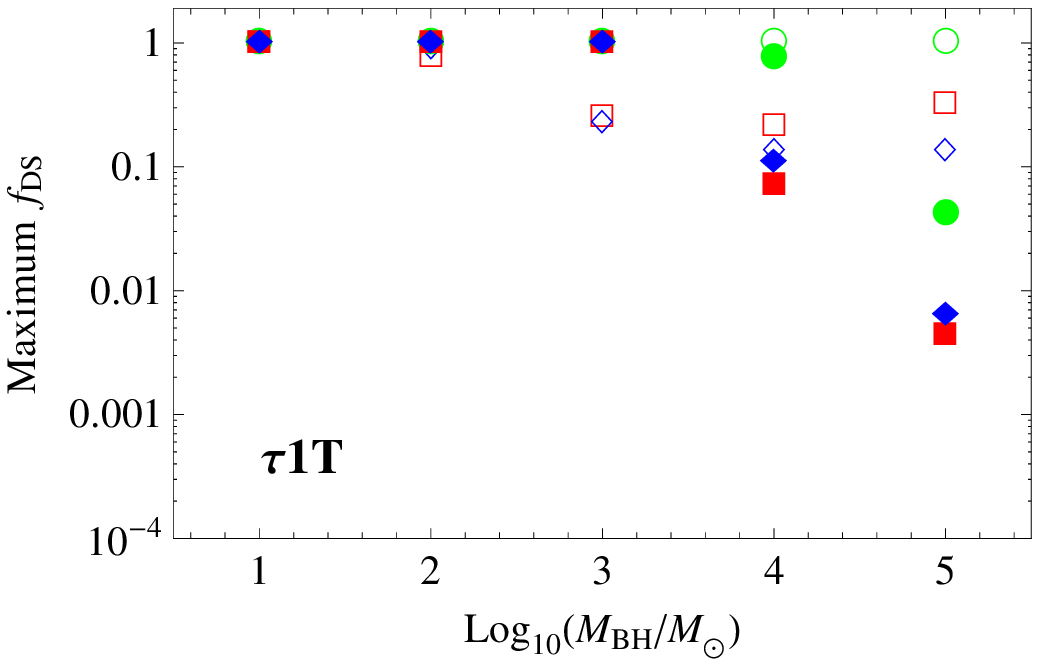,width=.4\textwidth}}\vspace{2mm}
\mbox{\epsfig{file=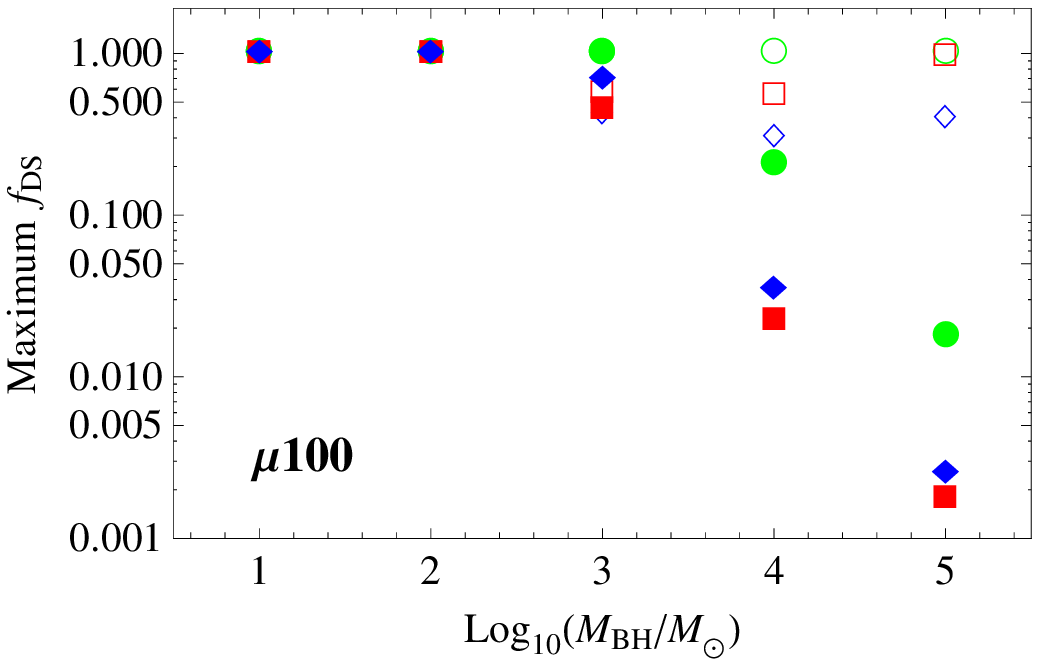,width=.4\textwidth}}\hspace{10mm}
\mbox{\epsfig{file=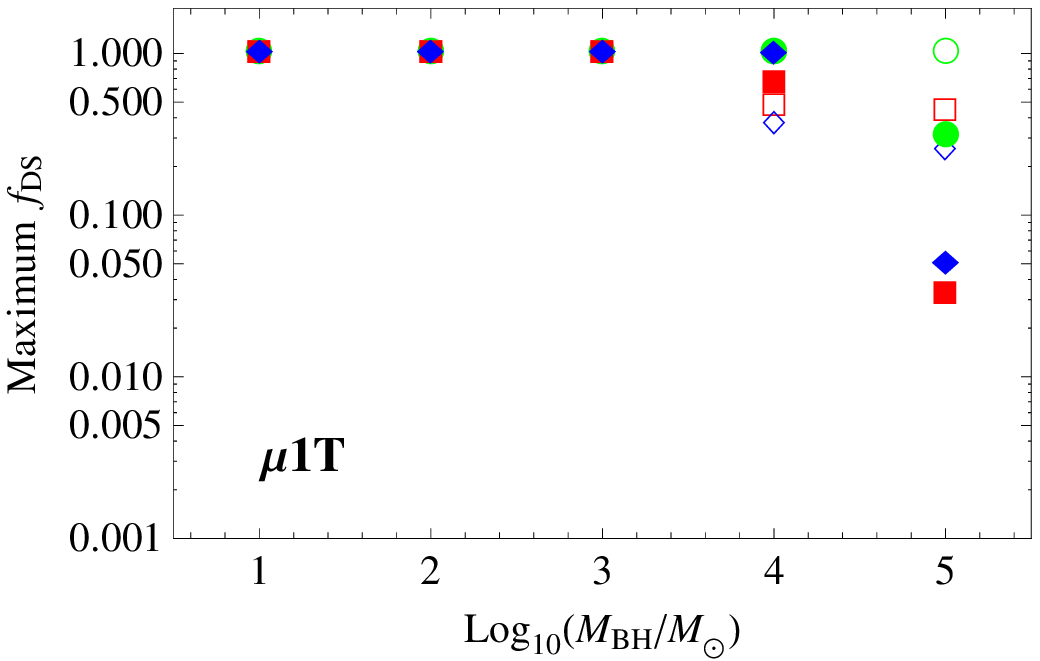,width=.4\textwidth}}
\end{center}
\caption{\it Maximum $f_{DS}$ as in Figs.~\ref{fig:fDSmax} and~\ref{fig:fDSmax5}, excluding the inner 5 kpc from the Galactic center and with the maximum $f_{DS}$ from point source brightness determined by the brightness of the brightest unassociated point source. The maximum $f_{DS}$ from the diffuse gamma-ray flux is as in Fig.~\ref{fig:fDSmax5}.  The range of $f_{DS}$ displayed is different than in previous figures for $\mu$ and $\tau$ final states.
\label{fig:fDSmax5u}}
\end{figure}

Finally, we return to the issue of the maximum possible brightness of DM spikes.  In the preceeding analysis, we consider only the most conservative criterion for the maximum DM spike brightness, namely that the brightest DM spike must not be brighter than the brightest FGST point source, Vela.  Given the size of Vela in the sky, it is highly unlikely that the brightest DM spike happens to lie along our line of sight to Vela.  In fact, even if it does, it is then extremely unlikely that the second brightest DM spike would also lie along our line of sight to another bright associated FGST source (and so on for subsequently dimmer DM spikes).  It is much more likely that the brightest DM spike does not lie precisely along our line of sight to Vela, or any other of the brightest FGST point sources.

Given the unlikelihood that DM spikes are hiding along our line of sight to the brightest {\it associated} FGST point sources, here we explore the possibility that the brightest DM spike must not be brighter than the brightest {\it unassociated} point source, which has an energy-integrated gamma-ray flux of $5.78\times 10^{-8}$ cm$^{-2}$s$^{-1}$, approximately $1/22$ that of Vela.  In Fig.~\ref{fig:fDSmax5u}, we present the maximum value of $f_{DS}$ as determined by Eq.~\ref{eq:PSconstraint} with the more stringent maximal flux.  As in Fig.~\ref{fig:fDSmax5}, we exclude any spikes located within 5 kpc of the Galactic center. Note that the ranges of $f_{DS}$ displayed for $\mu$ and $\tau$ final states are different from those in Figs.~\ref{fig:fDSmax} and~\ref{fig:fDSmax5}, and that there has been no change to the constraint derived from the diffuse gamma-ray flux.  It is clear that the nearest point source must now be even farther away, leading to much stronger constraints on $f_{DS}$.  Again, the limits on $f_{DS}$ are strongest for the most luminous individual spike scenarios (low WIMP mass, annihilations to $b\bar{b}$ or $W^+W^-$), and for the star formation histories that result in the largest number of DM spikes near our Solar System.  For annihilations of 100 GeV WIMPs to $b\bar{b}$ or $W^+W^-$, there are significant limits on $f_{DS}$ even for $100 M_\odot$ black holes, as would be expected from standard Population III.1 stars in the absence of a DS phase. 

The limits on $f_{DS}$ presented here have all been obtained assuming, as discussed in Sec.~\ref{sec:PS}, that the flux from an individual spike necessary for it to have been identified as an FGST point source does not depend on the location of the spike in the sky.  In fact, the flux required for a point source to have been identified does depend on the diffuse gamma-ray emission near the spike.  However, since some spikes would have been easier to identify while others may be in regions of the sky that inhibit identification, we estimate that our results are not sensitive to deviations from the average flux we've used. Specifically, since $D_{min}^{PS}$ only depends on the flux from the brightest point source, the constraint from near point sources is not affected at all, while the constraint from the diffuse flux comes from an ensemble of many sources, and will therefore be minimally affected.

If, however, the number of events required to identify a point source is significantly larger or smaller than the 50 per year that we have assumed, our results would change somewhat. In either case, there is no difference in the constraint arising from the brightest spike. If the number of events required is larger than 50 per year, more spikes would not be bright enough to have been identified as point sources, so the diffuse gamma-ray flux from all faint (non-point source) spikes would be larger than that presented here, and therefore the constraint on $f_{DS}$ arising from the diffuse flux would be {\it stronger} than what is presented in Figs.~\ref{fig:fDSmax}-\ref{fig:fDSmax5u}. Conversely, if sources producing fewer than 50 events per year in FGST have been identified as point sources, then fewer spikes would contribute to the diffuse flux, and the constraint on $f_{DS}$ arising from the diffuse flux would be {\it weaker} than what is presented in Figs.~\ref{fig:fDSmax}-\ref{fig:fDSmax5u}.
Again, for the calculation of the diffuse gamma-ray flux we include only spikes that result in less than 20 events per year above 1 GeV.
 If more (fewer) spikes actually contribute to the diffuse flux, then our constraint on $f_{DS}$ from the diffuse flux would improve (weaken), while our constraint from nearby point sources would be unaffected\footnote{We would also expect that there would be fewer (more) spikes that should have appeared as point sources in the First Source Catalog, and therefore fewer (more) spikes in Table~\ref{tab:nPS}.}.


\section{Antimatter and Neutrino Signatures of Nearby DM Spikes}
\label{sec:epnu}

In general, dark matter annihilations result in other potentially-detectable products in addition to photons.  For example, neutrinos, electrons and positrons, and a host of other Standard Model final states may be produced, depending on the annihilation mode. Neutrinos and antimatter present very interesting and potentially fruitful avenues in which to search for dark matter annihilations in nearby structures. While there are currently only limits on the dark matter annihilation rate to neutrinos and antiprotons, several recent experiments sensitive to cosmic ray electrons and positrons have found anomalous signals.

\subsection{Positrons: Comparison with PAMELA data}
\label{sec:positrons} 

The Payload for Antimatter Matter Exploration and Light-Nuclei Astrophysics (PAMELA) has observed an excess of cosmic ray positrons (relative to electrons) between 10 and 100 GeV~\cite{adriani}, confirming the hints from earlier experiments such as the High Energy Antimatter Telescope (HEAT)~\cite{heat} and the AntiMatter Spectrometer (AMS-01)~\cite{ams01}.  A surplus of cosmic ray electrons and/or positrons has also been confirmed by FGST~\cite{fgstelectrons}, however the spectral feature is less pronounced than that previously reported by the Advanced Thin Ionization Calorimeter (ATIC)~\cite{chang}. 

Although an excess of cosmic ray positrons may come from more conventional astrophysical sources such as pulsars~\cite{pulsars}, there has been a great deal of interest in the possibility that these signals might be a consequence of dark matter particles annihilating in the local halo of the Milky Way. Efforts to produce such signals with dark matter, however, have faced two major challenges. First, a very large annihilation rate is required, much larger than that expected from a smoothly-distributed thermal relic annihilating in the Milky Way halo.  Second, given the large annihilation rate, for most annihilation modes a positron excess would be accompanied by an excess of antiprotons.  However no such excess of cosmic ray antiprotons has been observed.

These two challenges have been addressed in a variety of ways, including string theory motivated supersymmetric models with non-thermal wino dark matter, as proposed by~\cite{kaneranwatson}.  Of particular relevance to our current work  is Ref.~\cite{nearbyclump}, which suggests that annihilations in a nearby clump of dark matter may be responsible for the positron excess. It may be extremely
unlikely that a DM subhalo with the required luminosity and distance
exists~\cite{vl2,Brun2009}, but the required luminosities and
distances do lie within the range of DM spike scenarios considered
here and in~\cite{bzs,bftz,tabp,bls}.
For example, an 800 GeV WIMP annihilating to $W^+W^-$ in the DM spike surrounding a $10^2 M_\odot$ black hole has a luminosity of $\sim 10^{37} \gamma/$s, according to Fig.~\ref{fig:IntLum}. It was found by Ref.~\cite{nearbyclump} that this is roughly the luminosity required to explain the PAMELA positron excess if the dark matter clump or spike is located approximately 1 kpc from our Solar System.  From the left panel of our Fig.~\ref{fig:PointSourceDist}, we see that for a $10^2 M_\odot$ black hole and annihilations to $W^+W^-$, the nearest spike as constrained by the FGST point source data may be as close as $\sim1/2$ kpc, potentially over-producing cosmic ray positrons. If that DM spike is 1 kpc from our Solar System, in agreement with PAMELA, it may already appear in the FGST catalog. 

The feasibility of explaining the excess of cosmic ray positrons with nearby DM spikes surrounding $10^5 M_\odot$ black 
holes was also addressed in Ref.~\cite{bls}, where it was confirmed that a DM spike $\sim1$ kpc from our Solar System could indeed reproduce the PAMELA excess.  Moreover, it was shown that a range of annihilation rates and spike distances could explain the PAMELA excess without overproducing the combination of electrons $+$ positrons observed by FGST. The implications of the existence of more than one nearby DM spike may be important as measurements of the flux of cosmic ray electrons and positrons improve.
A more detailed analysis of the interplay between gamma-ray and antimatter constraints in these scenarios, including the effects of multiple spikes according to the local distribution as predicted by VL-II, we leave for future work~\cite{PAMELAwip}.

\subsection{Neutrinos}
\label{sec:neutrinos}

Although the photon flux generally imposes a stronger constraint on dark matter annihilations, the neutrino flux may become more significant if there is a substantial branching fraction to leptonic final states. Searches for a diffuse neutrino flux from dark matter annihilations are limited by the appreciable background of atmospheric neutrinos above $\sim$ 40 MeV, though it has been shown that if there is a substantial branching fraction to leptonic final states, 
the neutrino flux may be a very useful tool to learn about the properties of dark matter through its annihilations in 
the Milky Way halo~\cite{yhba,ICgc, ICgh} and in the dark matter halos of Milky Way dwarf satellite 
galaxies~\cite{ICdwarfs,EssigSegue}.

The prospects for detecting neutrinos from dark matter annihilations in the spikes surrounding $10^5 M_\odot$ black holes were investigated in Ref.~\cite{zentnerNu}, where the number of DM spikes producing sufficient event rates at several neutrino telescopes was calculated. Ref.~\cite{zentnerNu} concludes that, based on the distribution of such black holes from~\cite{bzs}, neutrinos from $10^5 M_\odot$ black holes could be detected by both the IceCube neutrino detector at the South Pole and ANTARES, in the Mediterranean Sea. 

The most stringent limits on the flux from point sources in the Northern Hemisphere have been set by the AMANDA-II Collaboration~\cite{amandaPS}, though these limits apply only to neutrinos with $E_\nu >1.6$ TeV, a threshold too large to be relevant for the models considered here. The full IceCube/DeepCore neutrino detector will have improved sensitivity for point source searches with neutrino energies potentially as low as $\sim10$ GeV~\cite{icecubePS}.  The ANTARES neutrino telescope claims the best limits on the neutrino flux from point sources in the Southern Hemisphere for 10 GeV $< E_\nu <$ 100 TeV of $E_\nu^2 d\Phi/dE_\nu <$ few $\times 10^{-7}$ GeV$\,$cm$^{-2}\,$s$^{-1}$~\cite{antaresPS}.

\begin{figure}[h]
\begin{center}
\mbox{\epsfig{file=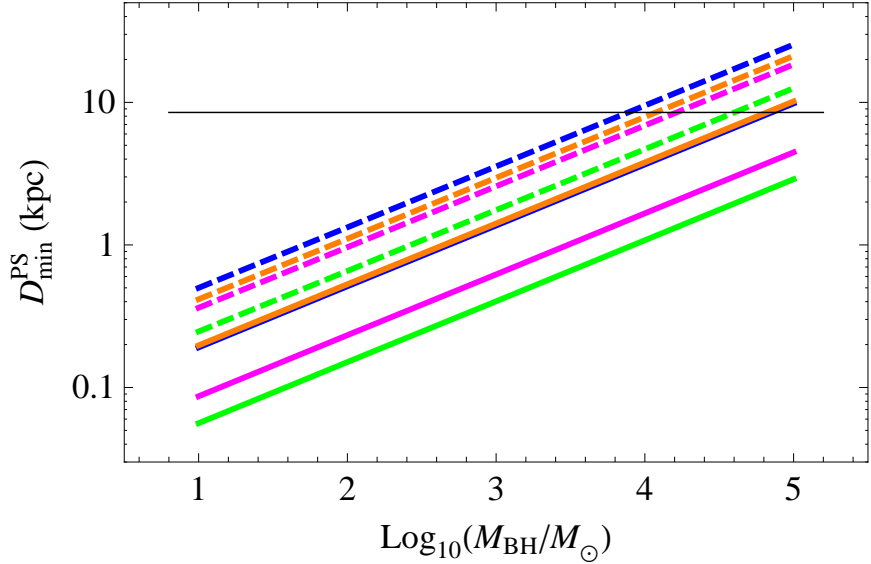,width=.7\textwidth}}
\end{center}
\caption{\it The minimum distance at which a spike may be found such that the flux of neutrinos from dark matter annihilation in the spike does not exceed the brightest neutrino point source as measured by Super Kamiokande.  This distance is plotted as a function of WIMP mass for Intermediate $z_f$.  From top to bottom, the contours are for models b100 (blue dashed), W100 (orange dashed), $\tau$100 (magenta dashed), $\mu$100 (green dashed), W1T (orange solid), b1T (blue solid), $\tau$1T (magenta), and $\mu$1T (green solid). The black horizontal line
indicates our distance from the GC.
\label{fig:DminNuMu}}
\end{figure}

Super-Kamiokande (SK), located in the Mozumi Mine in Japan, is sensitive to neutrinos from the Southern Hemisphere down 
to $E_\nu = 4.5$ MeV, where the flux of solar neutrinos dominates. Recently, SK performed a search for neutrino point 
sources and found that $\Phi \lesssim 10^{-7}$ cm$^{-2}$s$^{-1}$ for $E_\nu > 1.6$ GeV at declinations between 
$-90^\circ$ and $\sim 50^\circ$~\cite{superkPS}. We plot in Fig.~\ref{fig:DminNuMu} the minimum distance at which a DM 
spike must be located in order not to exceed this flux, assuming the source is located in the field of view of SK, as 
in Fig.~\ref{fig:PointSourceDist}.  Here we use full three flavor vacuum oscillations~\cite{pdg}, and consider only 
neutrinos that arrive at Earth as $\bar{\nu}_\mu$, as in Ref.~\cite{superkPS}. Comparing Fig.~\ref{fig:DminNuMu} and 
the left panel of Fig.~\ref{fig:PointSourceDist}, we see that neutrino point source brightness provides a slightly 
stronger constraint on the minimal distance of the nearest spike in all scenarios shown if the brightest DM spike must 
not be brighter than Vela in gamma-rays.  

Of course, unjustified assumptions have been made here; most obviously that the SK limit on the flux from point sources applies to the full sky.  There are many caveats, as well, such as the fact that the limit was placed under the assumption that all point sources emit neutrinos with a spectral index of $\gamma =2$, such that $\Phi_\nu \propto \Phi_\nu^{(0)} E^{-\gamma}$.  For softer spectra (larger $\gamma$), the limits are considerably weaker~\cite{superkPS}.  For these reasons we abandon, for now, the possibility of using neutrinos to constrain dark matter annihilations in density spikes surrounding $10-10^5 M_\odot$ black holes. We note, however, that future neutrino observations with the DeepCore supplement to the IceCube detector, which will probe energies down to $E_\nu \sim 10$ GeV (possibly with full-sky coverage)~\cite{deepcore}, as well as future observations from SK, may provide possibilities for novel signatures of nearby DM spikes in neutrinos.  The interplay of neutrino and gamma-ray astronomy in the indirect detection and identification of dark matter may be very important in the near future.

\section{Discussion and Conclusions}
\label{sec:conclusions}

In this paper we have examined $10-10^5 \msun$ black holes with DM spikes that formed in 
early minihalos and still exist in our Milky Way Galaxy today.  
Such black holes may arise as the 
remnants of Dark Stars after the dark matter fuel is exhausted and thermonuclear burning runs its course. 
Since the redshift at which the increasing UV background and/or metal enrichment results in the truncation
of Pop.~III.1 star formation is poorly constrained, we have examined three scenarios: Early
($z_f\approx23$), Intermediate ($z_f\approx 15$), and Late ($z_f\approx 11$).   We determined the $z=0$ distribution of DM spikes throughout the Galactic
halo from the Via Lactea II cosmological N-body simulation \cite{vl2}.  We find 409, 7983, and 12416 DM spikes in the Milky Way for the Early, Intermediate, and Late scenarios, respectively, for the 
case where  $f_{DS}=1$ (every minihalo hosts a BH with a spike, no mergers). 
 As the DM spikes around the black holes are the sites of significant DM annihilation, we examined the signatures of the spikes in gamma-rays, and commented on searches in $e^+/e^-$
and neutrinos.

The focus of this paper is the gamma-rays
from DM annihilation in the spikes around black holes in the Galaxy, and comparison with data from FGST.  We find that some or all
of the unassociated point sources observed by FGST could be black holes with DM spikes.  Indeed,
Table 2 shows the estimated number of point sources in the Milky Way halo for a variety of DM models, and one can see that in most cases more than the 368 observed sources are present. 
It is exciting to imagine that some significant fraction of the point sources might be due to DM annihilation 
near black holes in our Galaxy.
For further discussion of DM point sources in the FGST First Source Catalog, see Ref.~\cite{buckleyhooper}.

Additionally, one can use the FGST observations to place limits on the properties of DM
spikes around black holes.  We compare the gamma-ray flux from our models to both that from point sources and the diffuse
flux observed by FGST, and ensure that they do not produce gamma-rays
in excess of what is observed. We have found the maximum fraction of Pop.~III.1 star-forming minihalos that could have contained DSs and their black hole remnants.
Our results are shown in Figs.~\ref{fig:fDSmax}-\ref{fig:fDSmax5u}.  In general, it is clear that the
bounds are the strongest for the largest black hole masses and if star formation persisted to low redshift (i.e. the Late model where star formation ends at  $z_f=11$).  We have
also found  that our bounds are robust with respect to uncertainties related to dynamics near the Galactic center,
in that neglecting all DM spikes in the inner 5 kpc of the Galaxy does not substantially
affect the limits.  

 We note that our limits depend very sensitively on the DM annihilation channel and
 on the black hole mass. 
  One can see that even the conservative constraints from bright point sources, shown in Figs.~\ref{fig:fDSmax} and~\ref{fig:fDSmax5}, are stronger than those from the diffuse flux in the
case of annihilation primarily to $b \bar{b}$ or $W^+ W^-$.  These cases are most
typical of Minimal Standard Supersymmetric Model (MSSM) neutralinos.   The bounds
become more restrictive for higher mass black holes.  For example, the upper left panels in both
Figs.~\ref{fig:fDSmax} and \ref{fig:fDSmax5}
illustrate the bounds for 100 GeV WIMPs with annihilation primarily to $b \bar{b}$.
There is essentially no constraint for $m_{BH}\lesssim 100 \msun$, while the constraint on the fraction
of minihalos containing black hole spikes becomes $f_{DS} B_{b \bar{b}} < 1/10$ for $1000 \msun$ black holes for Intermediate and Late $z_f$.
For the case of 100 GeV WIMPs
and even more massive $10^5 \msun$ black holes, the bound becomes $f_{DS} B_{b \bar{b}} < .01$, regardless of $z_f$.  If the
number of DSs with these properties is low, it becomes more difficult to find
them in upcoming observations
with James Webb Space Telescope (JWST).  On the other hand,
for TeV-mass WIMPs,  the upper right hand panel of Fig.~\ref{fig:fDSmax5} shows a weaker
constraint for the $10^5 \msun$ black holes of $f_{DS} B_{b \bar{b}} < 1/10$.  Thus one out
of every ten minihalos with the right properties for Pop.~III.1 star formation might have hosted a DS with these characteristics.  This is a  good case for discovery of DS in JWST \cite{Freese:2010re}. 
From the remaining panels in Figs.~\ref{fig:fDSmax} and \ref{fig:fDSmax5}, one can see that 
the constraints are relatively weak for WIMPs that annihilate to leptonic final states.

While scenarios involving the larger black holes considered here are decidedly constrained for all models in Figs.~\ref{fig:fDSmax} and \ref{fig:fDSmax5}, it is remarkable that we are able to find conservative limits on $f_{DS}$ even for $100 M_\odot$ black holes for some dark matter models.  Black holes of this size are roughly what is expected from standard Pop.~III.1 star formation, if there is no DS phase.  We also point out that the smallest black holes we consider here, with $m_{BH}=10 M_\odot$, are actually disfavored by both theory~\cite{larson1999} and simulations~\cite{bcl1999,massivepop3sims} as the remnants of Pop.~III.1 stars, which are expected to be considerably larger. 

In addition to considering the most conservative case that allows for the possibility that the brightest DM spike could be located along our line of sight to the brightest FGST point source, we also consider the more likely scenario that the brightest DM spike is in fact not hiding along our line of sight to any of the brightest sources in the FGST catalog that are associated with astrophysical sources in other wavelengths.  If the brightest DM spike must not be brighter than the brightest unassociated object in the FGST First Source Catalog, $f_{DS}$ must indeed be quite small for $m_{BH}\gtrsim 1000 M_\odot$ for most dark matter models explored.  For annihilations of 100 GeV WIMPs to $b\bar{b}$ or $W^+W^-$, the limits on $f_{DS}$ are quite significant even for $100 M_\odot$ black holes, as would be expected from standard Population III.1 stars in the complete absence of a DS phase. In order to trust these stronger limits, however, we must be completely confident in our understanding of the brightest associated FGST point sources and confident that there is no DM spike hiding along the line of sight to any of them.

We stress that the limits derived here on $f_{DS}$ are related to the initial fraction of minihalos in which Population III.1 stars formed and the fraction of DM spikes subsequently disrupted by mergers as $f_{DS}=f_{DS}^0 (1-f_{merged})$.  While our estimates show that mergers don't change the number of DM spikes in our Galaxy by more than a factor of 2, a detailed understanding of black hole mergers is required to
translate our constraints on $f_{DS}$ to constraints on the fraction of minihalos that were capable of hosting Population III.1 stars that actually did, $f_{DS}^0$.
This information could then be used to place an explicit limit on the Population III.1 star formation rate.  If/when the annihilation properties of particle dark matter become known, we may be provided with some interesting hints about the formation of the first stars. 

 Here we have restricted our work to black holes with $m_{BH} \leq 10^5 M_\odot$ and WIMP masses in the range 100 GeV - 2 TeV.  Larger DSs are indeed possible~\cite{Freese:2010re}, the most massive of which are expected to be sufficiently bright to be detected by the Hubble Space Telescope (HST)~\cite{zackrisson2}. In fact, Ref.~\cite{zackrisson2} uses HST data to constrain the formation rate and lifetime of $10^7 M_\odot$ SMDSs.  It would be interesting to directly compare the constraints derived from FGST data with those from HST.
 It would also be very interesting to examine the possibility of lighter $\mathcal{O}(10)$ GeV WIMPs.  On the one hand, the annihilation rate (which scales inversely with the WIMP mass) would be larger, so the sources would be brighter.  On the other hand, the photon energies would be lower
so it's possible that these light WIMP scenarios are already tightly constrainted by the EGRET-measured photon flux.
This would be an interesting case to study, particularly since current data from direct
detection experiments seem to favor this WIMP mass range; DAMA/LIBRA and CoGeNT
data may be consistent with such a value \cite{bernabei, collar, savage, hooper}, though recent reports may indicate incompatibility with XENON10 data~\cite{sorensen}.

We also briefly discussed the relevance of DM spikes in relation to the anomalous cosmic ray positron excess measured by PAMELA.  DM spikes seem to be consistent with the dark matter annihilation interpretation of the excess. Further study of the distribution of spikes in the Milky Way halo and compatibility with cosmic ray antimatter spectra is merited~\cite{PAMELAwip}.

In this paper, we have shown how dark matter structures that exist today carry information about the properties of the first stars in the Universe, and we have explored a few ways to learn about them through existing gamma-ray data from FGST.  We look forward to the day when the properties of particle dark matter reveal themselves, and the implications for Dark Stars and dark matter astronomy unfold.

\section*{Acknowledgements}

J.D.~is supported by the Swiss National Science Foundation.
K.F.~thanks the Department of Energy and the Michigan Center for Theoretical Physics for support, 
and the Aspen Center for Physics for hospitality during the course of this research.
K.F.~thanks Paul Shapiro for helpful conversations.
P.S.~is supported by the National Science Foundation under Grant No.~PHY-0455649, and thanks the Michigan Center for Theoretical Physics for providing a stimulating working environment during the Dark Stars Workshop. P.S.~also thanks Eiichiro Komatsu for helpful conversations. D.S.~is supported by the Department of Energy.

\end{document}